\documentclass{article}

\usepackage{authblk}
\usepackage{microtype}
\usepackage{color}
\usepackage{amsmath}
\usepackage{amsthm}
\usepackage{subfigure}
\usepackage{amssymb}
\usepackage{graphicx}
\usepackage{subfigure}
\usepackage[linesnumbered,ruled]{algorithm2e}
\usepackage{dsfont}
\usepackage{extarrows}
\usepackage{multirow}
\usepackage{diagbox}

\def\>{\ensuremath{\rangle}}
\def\<{\ensuremath{\langle}}
\def\ra{\ensuremath{\rightarrow}}

\def\h{\ensuremath{\mathcal{H}}}
\def\la{\ensuremath{\leftarrow}}

\newcommand {\tr} {{\mathrm{tr}}}

\newtheorem{prob}{Problem}
\newtheorem{thm}{Theorem}[section]
\newtheorem{cor}{Corollary}[section]
\newtheorem{lem}{Lemma}[section]
\newtheorem{defi}{Definition}[section]

\newtheorem{exam}{Example}[section]

\newtheorem{rem}{Remark}

\title{Quantum Privacy-Preserving Perceptron}

\author[1]{Shenggang Ying}
\author[1,2,3]{Mingsheng Ying}
\author[1]{Yuan Feng}
\affil[1]{University of Technology Sydney, Sydney, Australia}
\affil[2]{Institute of Software, Chinese Academy of Sciences, China}
\affil[3]{Tsinghua University, China}
\begin{document}
\maketitle

\begin{abstract}With the extensive applications of machine learning, the issue of private or sensitive data in the training examples becomes more and more serious: during the training process, personal information or habits may be disclosed to unexpected persons or organisations, which can cause serious privacy problems or even financial loss. In this paper, we present a quantum privacy-preserving algorithm for machine learning with perceptron. There are mainly two steps to protect original training examples. Firstly when checking the current classifier, quantum tests are employed to detect data user's possible dishonesty. Secondly when updating the current classifier, private random noise is used to protect the original data. The advantages of our algorithm are: (1) it protects training examples better than the known classical methods; (2) it requires no quantum database and thus is easy to implement.
\end{abstract}

\section{Introduction}
Privacy-preserving machine learning and data mining has been a very active research area during the past two decades, with various algorithms developed for privacy-preserving recommender systems \cite{Ricci2011}, logistic regression \cite{Chaudhuri2009}, decision tree learning \cite{AgrawalS2000}, association rule mining \cite{EvfimievskiSAG2004,VaidyaC2002}, information sharing across private databases \cite{AgrawalES2003}, just name a few.
In particular, they found lots of successful applications in the real world, for instance, banking, social media, storage, and supermarket shopping \cite{AgrawalS2000,Chaudhuri2009,EvfimievskiSAG2004}.

Quantum algorithms have been proved to be faster than their classical counterparts for several important tasks. For example, Grover algorithm \cite{Grover1996} can search in an unstructured database with a quadratic speedup over any classical algorithm. Based on it, a quantum algorithm for counting was proposed in \cite{BrassardHT1998}. Recently, quantum computing stepped into the area of machine learning. A quantum support vector machine was developed in \cite{RebentrostML2014} for big data classification. A quantum algorithm based on quantum random access memory \cite{GiovannettiLM2008qram} was presented in \cite{LloydMR2013} for supervised and unsupervised machine learning. In \cite{WiebeKS2014DL}, a quantum deep learning algorithm was devised by generalizing the Boltzmann machine. Employing Grover search algorithm, a quantum perceptron model was introduced in \cite{KapoorWS2016Perceptron} that is able to find training examples, which cannot be  classified by the current classifiers correctly, faster than classical methods. On the other hand, several quantum protocols have been proposed for protecting security and privacy in communication and computation, for instance quantum key distribution BB84 \cite{BennettB1984}, quantum private queries \cite{GiovannettiLM2008PQ}, quantum bit escrow\cite{AharonovTVY2000}, quantum bit commitment \cite{HardyK2004}, and coin tossing \cite{ColbeckK2006}. All these quantum security protocols has a common procedure: they first prepare quantum states on random bases, and finally perform measurements on the post-action states to detect dishonesty or cheat.

This paper aims to develop a quantum algorithm for privacy-preserving machine learning. As a starting point, we focus on perceptron \cite{Rosenblatt1957}, which is the simplest model of artificial neural networks. The distinctive features of our protocol are summarised as follows:
\begin{itemize}
  \item It preserves the privacy of the data provider better than the known classical methods. We propose two different ways to protect the private information of training examples in the two steps of the perceptron algorithm. In the first step (Step \textbf{S1} in~Sec~\ref{sec:per}), quantum tests are employed to detect if the data user was trying to extract information about training examples by performing quantum measurements. Once such a dishonesty is detected, the data provider will terminate the entire protocol to prevent further information leakage.
      In the second step (Step \textbf{S2} in~Sec~\ref{sec:per}), the data provider publishes a distorted training example for the data user to update classifiers, a technique widely used in classical algorithms. The key difference to those classical algorithms in this step is that the random noise is private, i.e., unknown to the data user. This makes the data user unable to recover the original data. Remarkably, numerical results show that our quantum protocol can tolerate much more noise, thus providing stronger privacy protection, than classical methods such as~\cite{AgrawalS2000}, while still returning the correct result.
  \item The quantum algorithms for searching, counting and machine learning \cite{BrassardHT1998,GiovannettiLM2008PQ,Grover1996,KapoorWS2016Perceptron,LloydMR2013,RebentrostML2014,WiebeKS2014DL} require a quantum database.
% Our protocol does quantum communications and computations on a classical database, and the training examples is stored in a classical database.
  But our protocol does not assume a quantum database. Instead, it only makes use of a classical one which contains the training examples.
  Furthermore, no entanglement is required, except in the testing round for the data provider to detect possible dishonest behaviors of the data user. Therefore, our protocol is much easier to implement.
  %\item
  %This is another reason for an easier implementation; in particular its robustness.
\end{itemize}

The structure of this paper is as follows. Section \ref{Sec:Preliminary} provides necessary preliminaries. In Section \ref{Sec:GeneralProtocol}, we present a quantum privacy-preserving data system for computing a linear function $f(x)$, which is the basis of the perceptron model. A theoretical analysis of the privacy level of this algorithm is given in Section \ref{Sec:PrivacyAnalysis}. Then in Section \ref{Sec:QPerceptron}, the data system is deployed to the perceptron model of machine learning. The correctness of the quantum privacy-preserving perceptron algorithm, some further privacy analysis and numerical experiments are also presented. A comparison with classical algorithms and other quantum algorithms for a similar purpose is given in Section \ref{Sec:Conclusion}. A more detailed description of our algorithm, proofs of theorems, and more numerical experiments and comparisons are showed in the Appendix.

\subsection{Related Works}
This paper is a continuation of our previous works \cite{YingYF2015,YingYF2017}. The differences between this paper and \cite{YingYF2015,YingYF2017} are as follows. First, the problem considered in this paper is machine learning with perceptron. In contrast, the problem dealt with in \cite{YingYF2015,YingYF2017} is association rule mining: \cite{YingYF2015} is for semi-honest agents on vertically partitioned databases, and \cite{YingYF2017} for dishonest agents on a centralized database. Second, the approaches are different. For mining association rules, quantum circuits are deployed to encode private information for semi-honest agents \cite{YingYF2015} and quantum tests are employed to detect cheat of dishonest agents \cite{YingYF2017}. But when learning a perceptron, classical training vectors have to be used to update the current classifier. This will cause privacy leakage. So the method in \cite{YingYF2015,YingYF2017} cannot be directly employed to preserve privacy in perceptron learning. Therefore, in this paper, a new quantum protocol is presented.

\section{Preliminary}\label{Sec:Preliminary}
We assume the readers have some basic background in quantum computation. For more details, we refer to \cite{NielsenC2010}. In this section, we very briefly recall the problem of privacy-preserving machine learning and the perceptron model, with the main aim being to fix notations.

\subsection{Privacy-Preserving Machine Learning}
Many privacy-preserving problems in machine learning and data mining such as association rule mining and decision tree learning \cite{AgrawalS2000,EvfimievskiSAG2004,VaidyaC2002} can be reduced to the following problem:
%compute the value $f(\vec{x})$ for a function $f$ and an input $\vec{x}$ without leaking the information of $\vec{x}$. A further requirement is to keep the function $f$ secret. Formally, it can be described as the following:
\begin{prob}\label{Prob:General}
    Let $f:\mathds{R}^k\ra\{0,1\}$ be a function privately held by Bob,
%    \begin{equation}\label{equation-1}
%        f:\mathds{R}^k\ra\{0,1\},
%    \end{equation}
    and $\vec{x}$ an input for $f$ privately held by Alice. How can Alice and Bob collaboratively compute $f(\vec{x})$ without the information about $\vec{x}$ leaked to Bob?
\end{prob}

\subsection{Perceptron}
Perceptron \cite{Rosenblatt1957} is a linear classifier and the simplest model of artificial neural networks. %It generates a linear function $f(x)$ to classify input $x$ based on a training set. The problem can be described as follows.
It solves the following problem based on a training set \cite{FreundS1999,Rosenblatt1957}:
\begin{prob}
    Let $D=\<(\vec{x}_1,c_1),\cdots,(\vec{x}_N,c_N)\>$ be a training set where each $\vec{x}_i =(x_{i,1},x_{i,2},\cdots, x_{i,k})\in \mathds{R}^k$ is a training vector and $c_i\in\{0,1\}$ is the class it belongs to. The problem is how to find a function/classifier
    \begin{equation}\label{Eq:Perceptron}
        f(\vec{x}) = \begin{cases}
            1 & \mbox{ if } \vec{w}\cdot \vec{x}+b> 0,\\
            0 & \mbox{ otherwise, } %\vec{w}\cdot \vec{x}+b<0,
        \end{cases}
    \end{equation} where $\vec{w}=(w_1,\cdots,w_k)\in \mathds{R}^k$ and $b\in \mathds{R}$,
    such that for every training vector $\vec{x}_i$, $f(\vec{x}_i)=c_i$.
\end{prob}

\begin{figure}
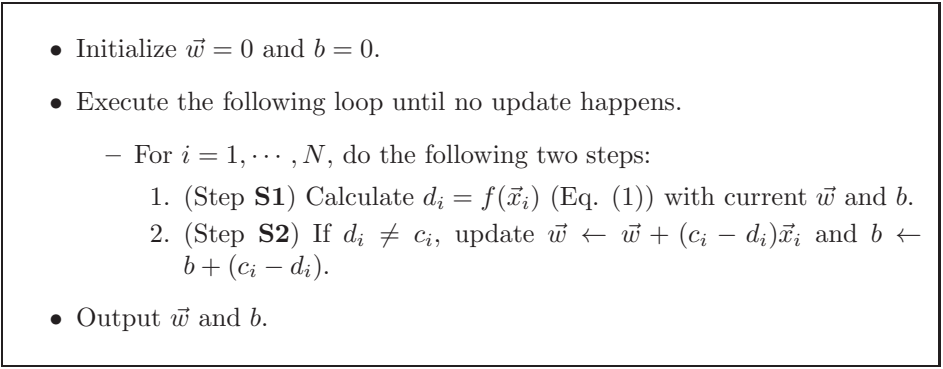

  \centering
  \fbox{\parbox{\textwidth}{
  \begin{itemize}
  \item Initialize $\vec{w}=0$ and $b=0$.
  \item Execute the following loop until no update happens.
  \begin{itemize}
    \item For $i=1,\cdots,N$, do the following two steps:
        \begin{enumerate}
            \item (Step \textbf{S1}) Calculate $d_i = f(\vec{x}_i)$ (Eq. \eqref{Eq:Perceptron}) with current $\vec{w}$ and $b$.
            \item (Step \textbf{S2}) If $d_i\neq c_i$, update $\vec{w} \la \vec{w}+(c_i-d_i)\vec{x}_i$ and $b \la b+(c_i-d_i)$.
        \end{enumerate}
  \end{itemize}
  \item Output $\vec{w}$ and $b$.
  \end{itemize}}
}
  \caption{The perceptron algorithm on a training set \cite{FreundS1999,Rosenblatt1957}.}\label{Fig:Perceptron}
\end{figure}

The perceptron algorithm dealing with this problem can be briefly summarized as in Figure \ref{Fig:Perceptron}. It is shown in \cite{Novikoff1963} that if such a $f$ exists, the original perceptron algorithm terminates up to a bounded number of updates (i.e., Step \textbf{S2}).

\subsection{Privacy Metric}\label{Sec:PrivacyMetric}
For the purpose of privacy analysis, a privacy metric is necessary. In this paper, we use the privacy metric defined in \cite{AgrawalS2000}. It can be formalized as follows:
\begin{defi}[Privacy metric \cite{AgrawalS2000}]\label{privacy-def}
    We say an attribute $x\in \mathds{R}$ in a training example has a privacy amount $l$ at $c\%$ confidence level, if $l$ is the minimum number such that the data user can estimate with $c\%$ confidence that $x\in[a,a+l]$ for some $a\in \mathds{R}$.
\end{defi}

In \cite{AgrawalS2000}, the privacy amount $l$ is only determined by the random noise added onto the original training examples. Suppose the noise $r$, which will be added onto $x$, is generated uniformly at random from $[-\delta,\delta]$. Then we say $x$ has a privacy amount $l=\frac{2c\delta}{100}$ with confidence $c\%$. For instance, if the confidence is $95\%$, i.e., $c=95$, the privacy amount is $1.9\delta$.

\begin{rem}Note for some special training examples, the privacy amount may decrease after the mining procedure. Suppose, $\vec{x}_j=(x_{j,1},x_{j,2}) = (20,5)$ is a training example of class 1, and uniformly distributed random noise $r\in [-5,5]$ is added onto $x_{j,1}$ to protect $x_{j,1}$. So generally $x_{j,1}$ has privacy amount 9.5 with confidence $95\%$. Now we further assume, after the mining or learning procedure, the data user finds that 99\% of training examples $\vec{x}_i$ of class 1 has the property $x_{i,1}>19$. Then the data user can conclude that $x_{j,1}>19$ with 99\% confidence. Thus the privacy amount should be $0.99(20+5-19)=5.94<9.5$.

Usually it is very hard to analyse the influence of the mining results on the privacy amount, since it depends on the specific problem. In \cite{AgrawalS2000}, this influence is not considered. So we do not consider this influence either. %Actually, we will find in our algorithm, since the data user does not know the detailed random generater,
\end{rem}

\section{A Quantum Privacy-Preserving Data System}\label{Sec:GeneralProtocol}
As stated in the introduction, the aim of this paper is to develop a quantum algorithm based on perceptron for machine learning. Our strategy is as follows: in this section, we present a quantum data system that protects privacy in computing a \textit{generic} linear function $f(\vec{x})$ in Problem \ref{Prob:General}.  Such a data system will be incorporated into the perceptron model for machine learning in Section 5.

\subsection{The Design Idea}
The basic idea for protecting Alice's privacy is similar to the common procedure used in \cite{AharonovTVY2000,BennettB1984,ColbeckK2006,GiovannettiLM2008PQ,HardyK2004}, which we mentioned in the introduction: Alice asks Bob to send back his state and then check it. But this idea is extended for our purpose in the following way: Alice asks Bob to perform three rounds of computation of $f$, one with the real data $\vec{x}$ and the other two with a randomly chosen test input. As the information about which round is for real computation and which rounds are merely for testing is kept secret from Bob, his attempt of attack (by performing measurement on the quantum systems received from Alice) will be detected by Alice with nonzero probability. It starts from the following simple observation.
\begin{exam}
    During the computation of $f(\vec{x})$, Alice sends the input $\vec{x}$ encoded as a quantum state $|\psi\>$ to Bob, and Bob computes the result on an ancilla qubit added after receiving $|\psi\>$. Alice may check whether Bob cheats if Bob is required to send back all qubits to Alice.  For instance, if Alice sends an entangled state $\frac{1}{\sqrt{2}}(|\vec{x}\>+|\vec{y}\>)$ with $\vec{x}\neq \vec{y}$ instead of the original data $|\vec{x}\>$ to Bob, then two situations arise after Bob performs the function $f$ on the states and sends them back to Alice:
            \begin{itemize}
              \item If Bob is honest, Alice receives $\frac{1}{\sqrt{2}}(|\vec{x}\>|f(\vec{x})\>+|\vec{y}\>|f(\vec{y})\>)$.
              \item If Bob is dishonest, that is he performs the projective measurement $\{|\vec{z}\> : \vec{z}\in \{0,1\}^k\}$ to learn Alice's input, he will obtain $\vec{x}$ or $\vec{y}$, each with probability $1/2$. However, since Bob does not know either $\vec{x}$ or $\vec{y}$ prior to measurements, he cannot make sure what indeed the measurement outcome is, and he cannot recover the expected state $\frac{1}{\sqrt{2}}(|\vec{x}\>|f(\vec{x})\>+|\vec{y}\>|f(\vec{y})\>)$. Thus Alice will be able to detect his dishonesty with probability 0.5.
            \end{itemize}
 %           Therefore, Alice can detect whether Bob is cheating by distinguishing these two situations.}
\end{exam}
Note that since Alice \textbf{does not} know either $f(\vec{x})$ or $f(\vec{y})$ before her final measurement,  she cannot distinguish these two situations directly. But we further observe that if Bob performs a measurement to read information, the final state is either $|\vec{x}\>|f(\vec{x})\>$ or $|\vec{y}\>|f(\vec{y})\>$ at random. In contrast, if Bob is honest, the final state is deterministic. This observation suggests that Alice should repeat the computation twice and compare the resulting states received back from Bob to determine whether Bob cheated.

\begin{rem} One may think that quantum private query \cite{GiovannettiLM2008PQ} can be used for our task here. But actually it is not appropriate because in this task, it requires (1) entanglement of all qubits, (2) quantum storages to store quantum states for a while, and (3) some public $\vec{x}_0$ with $f(\vec{x}_0)=0$ for any $f$. The first two requirements increase the difficulty for implementation, and usually the third cannot be satisfied in machine learning.
\end{rem}

\subsection{Notations in Our Protocol}
We now present our protocol by expanding the above basic idea. First, let us define some notations. As usual in quantum computing, to compute the function $f:\mathds{R}^k\ra\{0,1\}$  we assume a unitary operator $U_f$ with $U_f|0\>|\vec{x}\> =|0\>|\vec{x}\>$ and $U_f|1\>|\vec{x}\> =(-1)^{f(\vec{x})}|1\>|\vec{x}\>$ for any $\vec{x}\in \mathds{R}^k$. Indeed, $U_f$  can be seen as a $Z$ gate controlled by an oracle that computes $f$.
In this paper, the only useful input to $U_f$ is $|+\>|\vec{x}\>$, and $U_f$ transforms $|+\>$ to $|-\>$ if and only if $f(\vec{x})=1$:
\begin{equation}\label{Eq:Uf}
    U_f|+\>|\vec{x}\> = \begin{cases}
        |+\>|\vec{x}\> & \text{~if~} f(\vec{x})=0,\\
        |-\>|\vec{x}\> & \text{~if~} f(\vec{x})=1.
    \end{cases}
\end{equation}
The first qubit carrying $|+\>$ is called the result qubit, and the other qubits are called the data qubits.
%Remember from Eq. \eqref{Eq:Perceptron} that an input $\vec{x}\in \mathds{R}^k$.
%Remember that in Problem \ref{Prob:General} an input  $\vec{x}\in \mathds{R}^k$.
Suppose each component of $\vec{x}$ is represented by  an $n$-bit string. Then $\vec{x}\in \{0,1\}^{nk}.$

In order to detect Bob's attack, Alice needs to prepare a testing input state by the following steps:
\begin{enumerate}\item She generates a state
\begin{equation}\label{Eq:TestStateInitial}
    |\psi(\vec{y},0,0)\>=|+\>|\vec{y}\>,
\end{equation}
where $\vec{y}=y_1\cdots y_k\in\{0,1\}^{nk}$ is drawn uniformly at random;
\item She entangles the result qubit with the first data qubit by applying a CNOT gate on them;
\item With probability $1/2$, she applies a $Z$ gate on the result qubit, obtaining $|\psi(\vec{y},u,1)\>=\frac{1}{\sqrt{2}}(|0\>|y_1\>+(-1)^u|1\>|2^{n-1}\oplus y_1\>)|y_2\cdots y_k\>,$
%\begin{equation}
%    |\psi(\vec{y},u,1)\>=\frac{1}{\sqrt{2}}(|0\>|y_1\>+(-1)^u|1\>|2^{n-1}\oplus y_1\>)|y_2\cdots y_k\>,
%\end{equation}
where $u\in\{0,1\}$ is generated uniformly at random, and $\oplus$ denotes the addition modulo 2;
\item She randomly swaps the first data qubit and any other data qubit, thus obtaining
\begin{equation}\label{Eq:TestState}
    |\psi(\vec{y},u,m)\>=U_{SWAP(1,m)}|\psi(\vec{y},u,0)\>,
\end{equation}
where $m$ is chosen from $\{1,\cdots,kd\}$ uniformly at random.
It is worth noting that from Eqs. \eqref{Eq:Uf} and \eqref{Eq:TestStateInitial}, $|\psi(\vec{x},0,0)\>=|+\>|\vec{x}\>$ is a state for the computation. So, $|\psi(\vec{y},u,m)\>$ can be used to  represent either a test state ($m>0$) or a computational state ($m=0$).
\end{enumerate}

\subsection{System Description}
Now we can present our data system for computing $f(\vec{x}_i)$. In order to compute $f(\vec{x}_i)$, Bob needs to query the data system to get access to $\vec{x}_i$. Once Bob queries the data system, it answers three quantum states in sequence and requires Bob to send back each state after his action. The second (resp. third) state is sent once the first (resp. second) state is received back from Bob after his actions.

The procedure of this data system to answer $\vec{x}_i$ is given in Algorithm \ref{alg:ProtocolComputeF}. Algorithm \ref{alg:ProtocolComputeF} calls procedure \ref{alg:ProcedureCompute} three times, among which two are devoted to detecting Bob's attack and one to computing $f(\vec{x})$. In procedure \ref{alg:ProcedureCompute}, we use $Z_0$ to indicate the $Z$ gate on the result qubit. The controlled-NOT gate $U_\mathit{CNOT}$ is applied on the result qubit and the first data qubit, with the former being the control qubit.

\SetKwData{Left}{left}\SetKwData{This}{this}\SetKwData{Up}{up}
\SetKwFunction{Union}{Union}\SetKwFunction{FindCompress}{FindCompress}
\SetKwInOut{Input}{Input}\SetKwInOut{Output}{Output}
\SetKwInOut{Parameter}{Parameter}

\begin{algorithm}%[ht]
    \Input{$\vec{x}$}
    %\Parameter{$\vec{x}_i\in\{0,1\}^{nk}$}
    \Begin{
    Generates uniformly at random $i\in \{1,2,3\}$, $\vec{y}\in\{0,1\}^{nk}$, $u\in\{0,1\}$, and $m\in\{1,\cdots,nk\}$\;
    Consecutively executes the procedure \ref{alg:ProcedureCompute} three times, the $i$-th time with data-input $(\vec{x},0,0)$ and the other two  with test-input $(\vec{y},u,m)$\;
    If different results are obtained from the two test executions, then terminates the protocol\;\label{line:TM}
    Otherwise, sends the execution result of the procedure with data-input to Bob\;
    }
    \caption{Quantum privacy preserving data system to solve Problem~\ref{Prob:General}. Executed by Alice.}\label{alg:ProtocolComputeF}
\end{algorithm}

\begin{procedure}%[h]
    \Input{ $\vec{y}\in\{0,1\}^{nk},u\in\{0,1\},m\in\{1,\cdots,nk\}$}
 %   \Output{$f(\vec{y})$.}
    \Begin{
    Prepares $|\psi(\vec{y},u,m)\>$ in Eq.\eqref{Eq:TestState}\;
    Calls procedure \ref{alg:BobCompute} with parameter $|\psi(\vec{y},u,m)\>$, and let $|\varphi\>$ be the returned quantum state from Bob\;\label{line:CBob}
    Applies $U_{CNOT} Z_0^u U_{SWAP}(1,m)$ on $|\varphi\>$\;
    Measures all data qubits of $|\varphi\>$ with basis $\{|0\>,|1\>\}$.  If the outcome is not $\vec{y}$, terminates the entire protocol\;\label{line:CM}
    Measures the remaining qubit of $|\varphi\>$ with basis $\{|+\>,|-\>\}$. If the outcome is $|+\>$, then return 0; otherwise return 1\;
    }
    \caption{AliceCompute($\vec{y},u,m$). Executed by Alice.}\label{alg:ProcedureCompute}
\end{procedure}

\begin{procedure}%[h]
    \Input{ a state $|\psi\>$ (received from Alice) in $\h_R\otimes \h_D$}
 %   \Output{$f(\vec{y})$.}
    \Begin{
    Applies $U_f$ on $|\psi\>$ to obtain $|\varphi\> = U_f|\psi\>$\;\label{line:BobAttackMeasure}
    Returns (Sends back to Alice) $|\varphi\>$\;
    }
    \caption{BobCompute($|\psi\>$). (Executed by Bob)}\label{alg:BobCompute}
\end{procedure}

\section{Privacy Analysis for Algorithm \ref{alg:ProtocolComputeF}}\label{Sec:PrivacyAnalysis}
In this section, we show how privacy is preserved in Algorithm \ref{alg:ProtocolComputeF}. In order for a better understanding, let us start with two examples.

\subsection{Examples of Attacks}
Bob's attacks take place in Procedure \ref{alg:BobCompute}. He can perform measurements or use controlled-NOT gate and additional qubits to entangle the data qubits, instead of applying $U_f$ honestly.
\begin{exam}[Attack by measurements] Suppose Bob tries to get Alice's private information $\vec{x}$ by performing measurements on the data qubits.
That is, instead of following the protocol, he does the following two steps to replace Line~\ref{line:BobAttackMeasure} in procedure \ref{alg:BobCompute}:
{\rm
\begin{itemize}
    \item[\ref{line:BobAttackMeasure}.1] Measures all data qubits of $|\psi\>$ with basis $\{|0\>,|1\>\}$;
    \vspace{-0.5em}
    \item[\ref{line:BobAttackMeasure}.2] Records the measurement outcome $z\in \{0,1\}^{nk}$ and let $|\varphi\>$ be the post-measurement state of the system;
\end{itemize}
}
Unfortunately for him, he may attack on a test state $|\psi(\vec{y},u,m)\>$. We take $m=1$ and $u=0$ as an example. Then the test state is
$$|\psi(\vec{y},0,1)\>=\frac{1}{\sqrt{2}}(|0\>|y_1\>+|1\>|2^{n-1}\oplus y_1\>)|y_2\cdots y_k\>.$$
The state $|\varphi\>$ Alice receives
 will be $|0\>|y_1\>|y_2\cdots y_k\>$ or $|1\>|2^{n-1}\oplus y_1\>|y_2\cdots y_k\>$ with a fifty-fifty chance.
 Note that $U_{CNOT} Z_0^u U_{SWAP}(1,m)=U_{CNOT}$ (as $u=0$, $m=1$). Alice will get $|0\>|\vec{y}\>$ or $|1\>|\vec{y}\>$.
%\begin{equation*}
%    \begin{cases}
%        |0\>|y_1\>|y_2\cdots y_k\>\ra |0\>|\vec{y}\>,\\
%        |1\>|2^{n-1}\oplus y_1\>|y_2\cdots y_k\> \ra |1\>|\vec{y}\>.
%    \end{cases}
%\end{equation*}
In either case, the measurement outcome on the result qubit (Line~\ref{line:CM} in procedure~\ref{alg:ProcedureCompute}) will be $|+\>$ or $|-\>$ with equal probability. Thus, no matter what Bob does in the other calls of procedure \ref{alg:BobCompute}, the probability that $g_1\neq g_2$ is $0.5q+0.5(1-q)=0.5$, where $q$ is the probability that the outcome in the other call is $|+\>$. This means that the attack will be detected with probability 0.5. The general case for $m>1$ is similar.

One trick for Bob to pass the test is to guess $m$ and $u$. However, since $m$ and $u$ are generated uniformly at random, the probability that he guesses $m$ and $u$ correctly is only $\frac{1}{2nk}$. Another trick is to perform measurements only on part of the data qubits. This reduces the probability to be detected, but at the same time decreases the obtained information as well.
\end{exam}

\begin{exam}[Attack by entangling the data qubits to new ancilla qubits] A more sophisticated attack for Bob is to add $nk$ blank qubits, and use CNOT gate to copy information. However, he will still be detected if he unfortunately attacks on a test state. Again, take $m=1$ and $u=0$ as an example.
%
%to the end of qubits $|\psi(\vec{x},0,0)\>$,  use the CNOT gate to copy $\vec{x}$ and get $|+\>|\vec{x}\>|\vec{x}\>$. If Bob attacks on a test state, this can be detected. Without loss of generality, we assume $m=1$ and $u=0$.
%
The state in procedure \ref{alg:ProcedureCompute} with the input parameters $(\vec{y},0,1)$ evolves as follows:
\begin{enumerate}\item After Bob's attack, it becomes
$$\frac{1}{\sqrt{2}}(|0\>|\vec{y}\>|\vec{y}\>+|1\>|2^{n-1}\oplus y_1\>|y_2\cdots y_k\>|2^{n-1}\oplus y_1\>|y_2\cdots y_k\>).$$
%Then Bob sends the first $nk+1$ qubits back to Alice.
\item Alice performs $U_{CNOT}$ on the first $nk+1$ qubits and gets $$\frac{1}{\sqrt{2}}(|0\>|\vec{y}\>|\vec{y}\>+|1\>|\vec{y}\>|2^{n-1}\oplus y_1\>|y_2\cdots y_k\>).$$
%Note here that Alice only holds the first  $nk+1$ qubits of this state.
\item The measurement outcomes are $|+\>|\vec{y}\>$ or $|-\>|\vec{y}\>$ each with probability $1/2$.
\end{enumerate}
Similarly to the previous example, this attack will be detected with probability $1/2$.\end{exam}

\subsection{Theoretical Analysis}
In general, it is very hard to precisely define Bob's cheat behaviour because Alice is just a data system and does not know what Bob will do. On the other hard, since the training examples are stored in quantum state, dishonest Bob has to perform measurements to read out information. So, it is reasonable to define Bob's cheat as measurements, no matter these measurements are performed before or after he sending states back to Alice.
Then the theoretical privacy analysis of Algorithm \ref{alg:ProtocolComputeF} can be concluded as the following theorem. This theorem shows that Algorithm \ref{alg:ProtocolComputeF} achieves a kind of cheat-sensitive security, which is defined in \cite{HardyK2004}.
\begin{thm}\label{thm:SecurityA1}
    In Algorithm \ref{alg:ProtocolComputeF}, once Bob performs measurements (in his action, Procedure \ref{alg:BobCompute}) on Alice's state or entangles new qubits to this state, it will be detected with positive probability.
\end{thm}

The above theorem only gives a qualitative analysis.
%It is really hard to give a quantitative analysis, since Bob can choose different measurement bases and the probability is dependent on the chosen basis. The only clear thing is that the less information Bob gets from measurements, the less probability that it will be detected.
%For instance,
%\begin{itemize}
%  \item if Bob performs measurements with basis $\{|0\>,|1\>\}$ on all data qubits, it will be detected with probability 0.5;
%  \item if Bob performs measurements with measurement operators $M_0=\frac{1}{2}I-\epsilon |0\>\<0|,M_1=\sqrt{I-M_0^\dag M_0}$ with very small $\epsilon>0$, it will be detected with a very small probability.
%\end{itemize}
%From these two cases, we can find that
%There is another way to analyse privacy.
To analyse the algorithm quantitatively, we turn to compute the expected privacy amount of the entire database after the whole machine learning procedure. The privacy metric we employed here is that given in Definition \ref{privacy-def}. The confidence we choose in this paper is 95\%. In order to
estimate that an attribute $x_{i,j}$ is in some interval with confidence 95\%, Bob has to perform measurements on first qubits of $x_{i,j}$ with bases close to $\{|0\>,|1\>\}$. (If the confidence is chosen to be 100\%, the basis should exactly be $\{|0\>,|1\>\}$.) To simplify the presentation, we assume that Bob always performs measurements with basis $\{|0\>,|1\>\}$.
Then we have the following theorem.
\begin{thm}\label{thm:Probability}
    Assume in Algorithm \ref{alg:ProtocolComputeF} each attribute $x_{i,j}$ is represented by a $n$-bit binary string, where $n_1$ bits are before the decimal mark and $n-n_1$ are after. Suppose Bob performs measurements with basis $\{|0\>,|1\>\}$ on some qubits to reduce its privacy amount. Then
    \begin{enumerate}
      \item if the privacy amount of $x_{i,j}$ is reduced from $2^{n_1}$ to $2^{n_1-n_2}$ with confidence at least 95\%, Bob's cheat will be detected with probability $\frac{n_2-1}{2nk}$.
      \item if the privacy amount of each attribute of $\vec{x}_{i}$ is reduced to $2^{n_1-n_2}$ with confidence at least 95\%, Bob's cheat will be detected with probability  $\frac{n_2k-1}{2nk}$.
    \end{enumerate}
\end{thm}

%In the above theorem, we say the probability is approximately $\frac{n_2}{4nk}$ or $\frac{n_2}{4n}$, because one-bit information $f(\vec{x}_i)$ can be derived from the result of Algorithm \ref{alg:ProtocolComputeF} to reduce privacy amount by Bob. We can not exactly analyse the influence of  $f(\vec{x}_i)$ on the privacy amount, since it depends on the specific $\vec{x}_i$ and $f$. So here we roughly say this bit can provide one bit of information about $\vec{x}_i$.

\section{Quantum Privacy-Preserving Perceptron}\label{Sec:QPerceptron}
Our quantum privacy-preserving data system for computing $f(\vec{x})$ is presented in Section 3 and carefully analysed in Section 4. Now we use it to develop a quantum privacy-preserving perceptron algorithm.

\subsection{Sketch of Quantum Perceptron}\label{sec:per}
%Due to the limit of space, here we only present a sketch of the quantum perceptron. A detailed description is given as Algorithm \ref{alg:ProtocolPerceptron} in Appendix \ref{Apd:AlgorithmPerceptron}.

In this subsection, we give a brief description of our protocol. The original (classical) perceptron algorithm has only two steps, i.e. \textbf{S1} and \textbf{S2} in Figure \ref{Fig:Perceptron} involving the database (training set) $D$. So, if we want to expand it in order to preserve privacy, certain modifications should be introduced in these two steps.

Usually a classical protocol preserves privacy by adding noise into the original data: $\vec{x}'=\vec{x}+\vec{\tau}$, and the training set $D$ is distorted to $D'$. Then the data user reconstructs the original data distribution based on $D'$ and the way that $\tau$ is generated. Finally the learning algorithm is executed on the reconstructed distribution \cite{AgrawalS2000}.

The idea in our algorithm is quite different. The sketch of this protocol is presented in Fig. \ref{Fig:QPerceptron}, and the detailed algorithm is given as Algorithm \ref{alg:ProtocolPerceptron} in Appendix \ref{Apd:AlgorithmPerceptron}.
\begin{figure}
  \centering
  \fbox{\parbox{\textwidth}{
  \begin{itemize}
  \item Bob initializes $\vec{w}=0$ and $b=0$.
  \item Alice and Bob execute the following steps until no update happens.
  \begin{itemize}
    \item Alice secretly generates $u\in\{0,1\}^{\log N}$ uniformly at random.
    \item For $i=1,\cdots,N$, do the following two steps:
        \begin{enumerate}
            %\item Bob queries $i$-th training example from Alice.
            \item Alice secretly generates $\vec{r}$ by her private random number generator and then sends $c' = c_{i\oplus u}$ and $\vec{x}' = \vec{x}_{i\oplus u}+\vec{r}$ to Bob.
            \item (Step \textbf{S1}) Bob calculates $d = f(\vec{x}_{i\oplus u})$ with current $\vec{w}$ and $b$ on Alice's quantum data system, i.e., Algorithm \ref{alg:ProtocolComputeF}.
            \item (Step \textbf{S2}) If $d\neq c'$, Bob updates $\vec{w} \la \vec{w}+(c'-d)\vec{x}'$ and $b \la b+(c'-d)$.
        \end{enumerate}
  \end{itemize}
  \item Bob gets the final $\vec{w}$ and $b$.
  \end{itemize}}}
  \caption{The sketch of quantum privacy-preserving perceptron algorithm.}\label{Fig:QPerceptron}
\end{figure}
Here, we explain some key points in our protocol:
\begin{itemize}
  \item At the beginning of each outer loop, Alice permutes her database. This is done by first generating a private random number $u\in\{0,1\}^{\log N}$. Then each index $i$ is replaced by $i\oplus u$, the bit-by-bit XOR of $i$ and $u$.
  \item In the \textbf{S1} step, Algorithm \ref{alg:ProtocolComputeF} is employed on the permutated database based on the original database $D$ to preserve the original training data.
  \item In the \textbf{S2} step, a distorted training example $\vec{x}' = \vec{x}_{i\oplus u}+\vec{r}$ is published for updating $\vec{w}$ and $b$:
    \begin{equation}\label{Eq:UpdateH}
        \vec{w} \la \vec{w}+(c_{i\oplus u}-d)\vec{x}'~\text{and}~ b\la b+c_{i\oplus u}-d.
    \end{equation}
    In this step, $\vec{r}$ is generated by random number generator $R(\delta)$, where $\delta$ is a parameter. Here
    \begin{itemize}
      \item the details of $R(\delta)$ including $\delta$ are kept secret by Alice;
      \item the only requirement for $R$ and information known by Bob is that the mean value of random numbers generateor is 0.
    \end{itemize}
\end{itemize}

The privacy analysis will be given later in Section \ref{Sec:PrivacyAlice}. Here we first prove the correctness of our algorithm. Since a distorted training example is used to update $w$ and $b$, the correctness of our algorithm is not obvious. Fortunately, it can be proved in a way similar to \cite{Novikoff1963}.
\begin{thm}\label{thm:Correctness}
    Suppose (1) a random number generator of mean 0 is employed in Step \textbf{S2} in Algorithm \ref{alg:ProtocolPerceptron}, and (2) the original perceptron learning algorithm (Fig. \ref{Fig:Perceptron} or \cite{FreundS1999,Rosenblatt1957}) terminates. Then Algorithm \ref{alg:ProtocolPerceptron} will eventually terminate and generate a correct classifier on the same training set.
\end{thm}

This theorem implies that the correctness is independent of the choice of the random number generator. Therefore, Alice can keep the random number generator secret. This is very different from the classical case. In the classical machine learning situations (for instance \cite{AgrawalS2000}), it has to be published to Bob to guarantee the correctness of the final results.

To give a direct view and show the influence of different random number generator on our quantum protocol, numerical experiments are carried out with several different generators and shown in Table \ref{Table:Generator}.
\begin{table}
    \center
    \begin{tabular}{c|l}
      Generator & Description\\
      \hline\hline
      $R_0(\delta)$ & Uniform distribution from $[-\delta,\delta]$. \\\hline
      \multirow{2}{*}{$R_1(\delta)$} & Generate $r$ according to $R_0(\delta)$.\\
      & If $r>0$ then $r\la r+0.5\delta$, and if $r<0$ then $r\la r-0.5\delta$.\\\hline
      \multirow{2}{*}{$R_2(\delta)$} & Generate $r$ according to $R_0(\delta)$.\\
      & If $r>0$ then $r\la 2r$, and if $r<0$ then $r\la r-0.5\delta$. \\\hline
      $R_3(\delta)$ &Normal distribution with mean 0 and standard deviation $\delta$. \\\hline
      \multirow{2}{*}{$R_4(\delta)$} & Generate $r$ according to $R_3(\delta)$.\\
      &If $r>0$ then choose $r$ from $(0,1.5934\delta]$  uniformly randomly. \\
    \end{tabular}
    \caption{Random generators in numerical experiments. In practice, all the numbers are rounded to a precision of $1/1024$.}\label{Table:Generator}
\end{table}
%So it generates $r$ uniformly at random from the set discretized from $[-1.5\delta,-0.5\delta)\cup\{0\}\cup(0.5\delta,1.5\delta]$.
These experiments are done on three different training sets, which are generated in three different ways.
All the numerical results are averaged over 100 executions due to the randomness of our quantum protocol.
Our protocol succeeds with probability 1 for all different random number generators and all $\delta=\frac{1}{1024},\frac{1}{512},\cdots,128$.
%So we omit figures showing the correct probabilities, and only give figures showing the average rounds (outer loops).

\begin{figure}
  \centering
  \subfigure[Training set 1.]{\includegraphics[width=4cm]{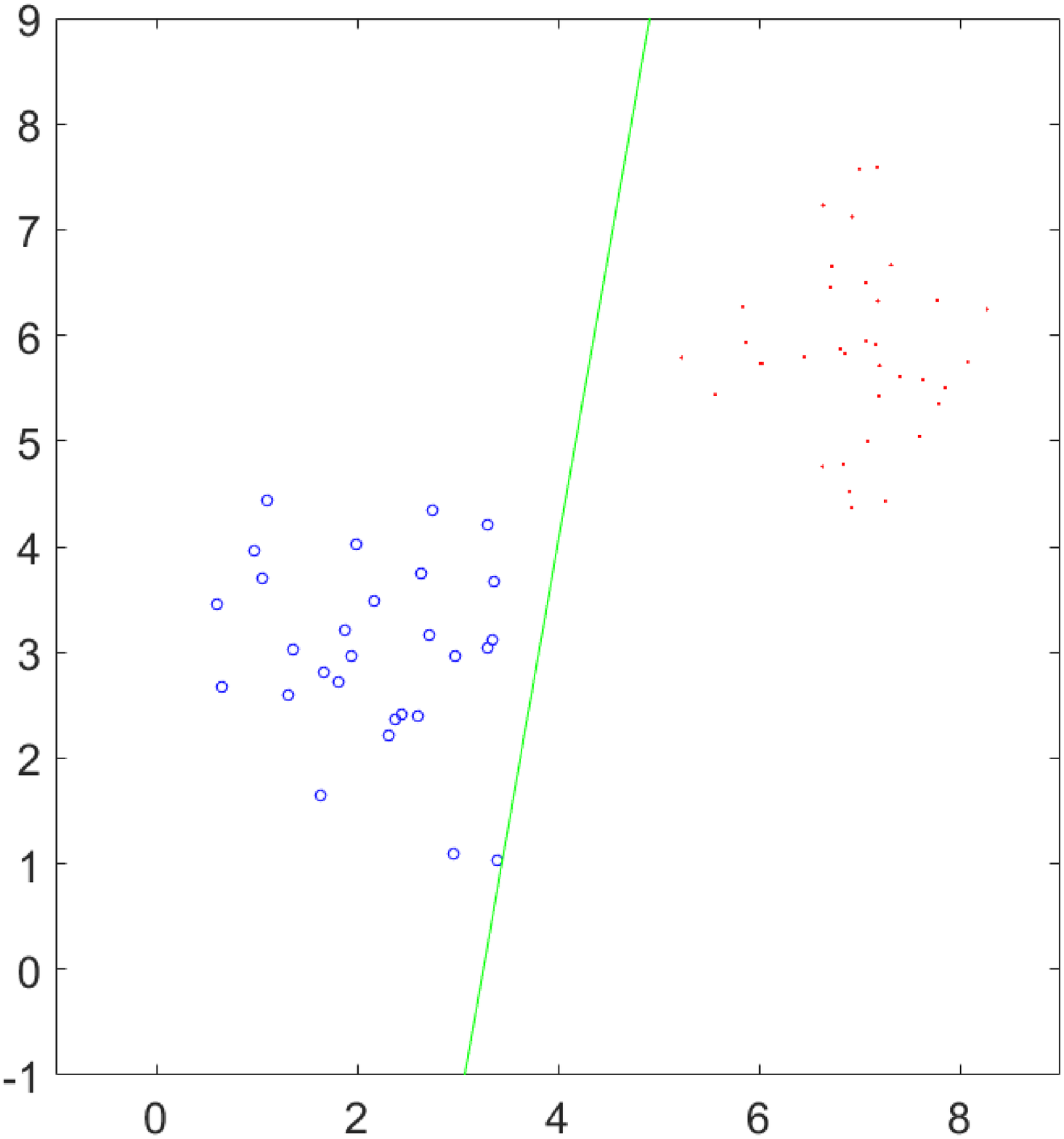}}
  \subfigure[Average running rounds on training set 1.]{\includegraphics[width=8cm]{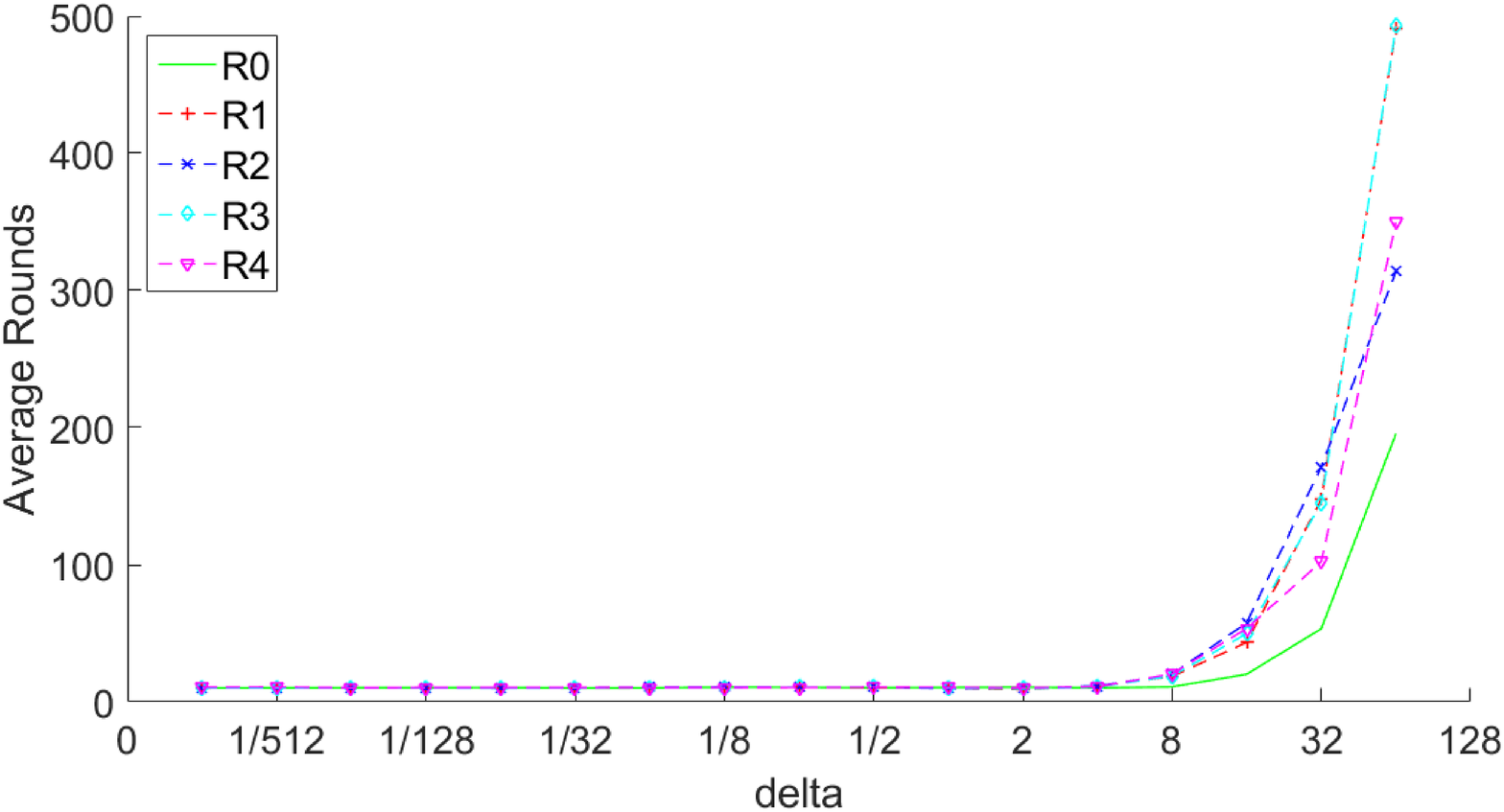}}
  \subfigure[Training set 2.]{\includegraphics[width=4cm]{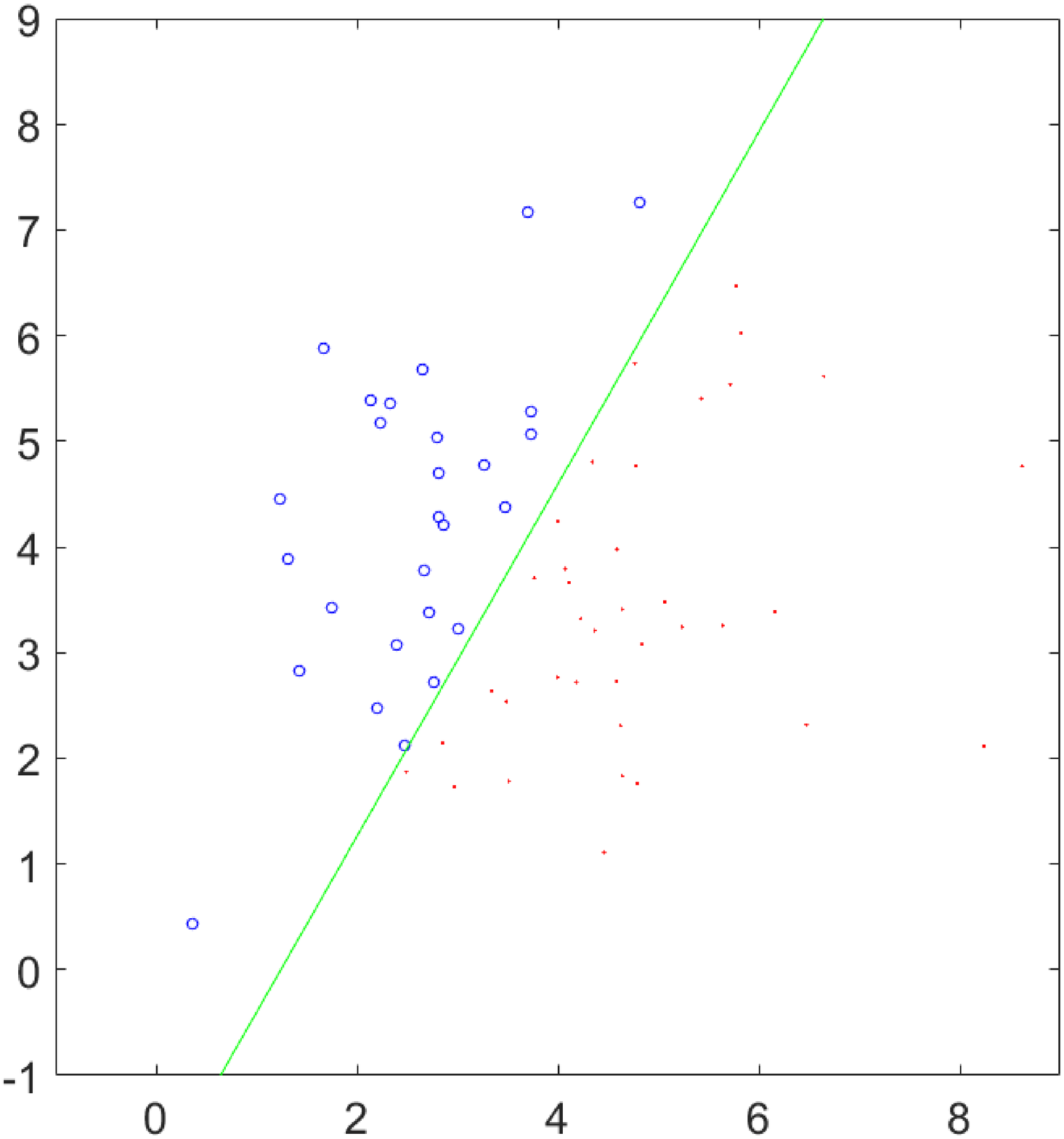}}
  \subfigure[Average running rounds on training set 2.]{\includegraphics[width=8cm]{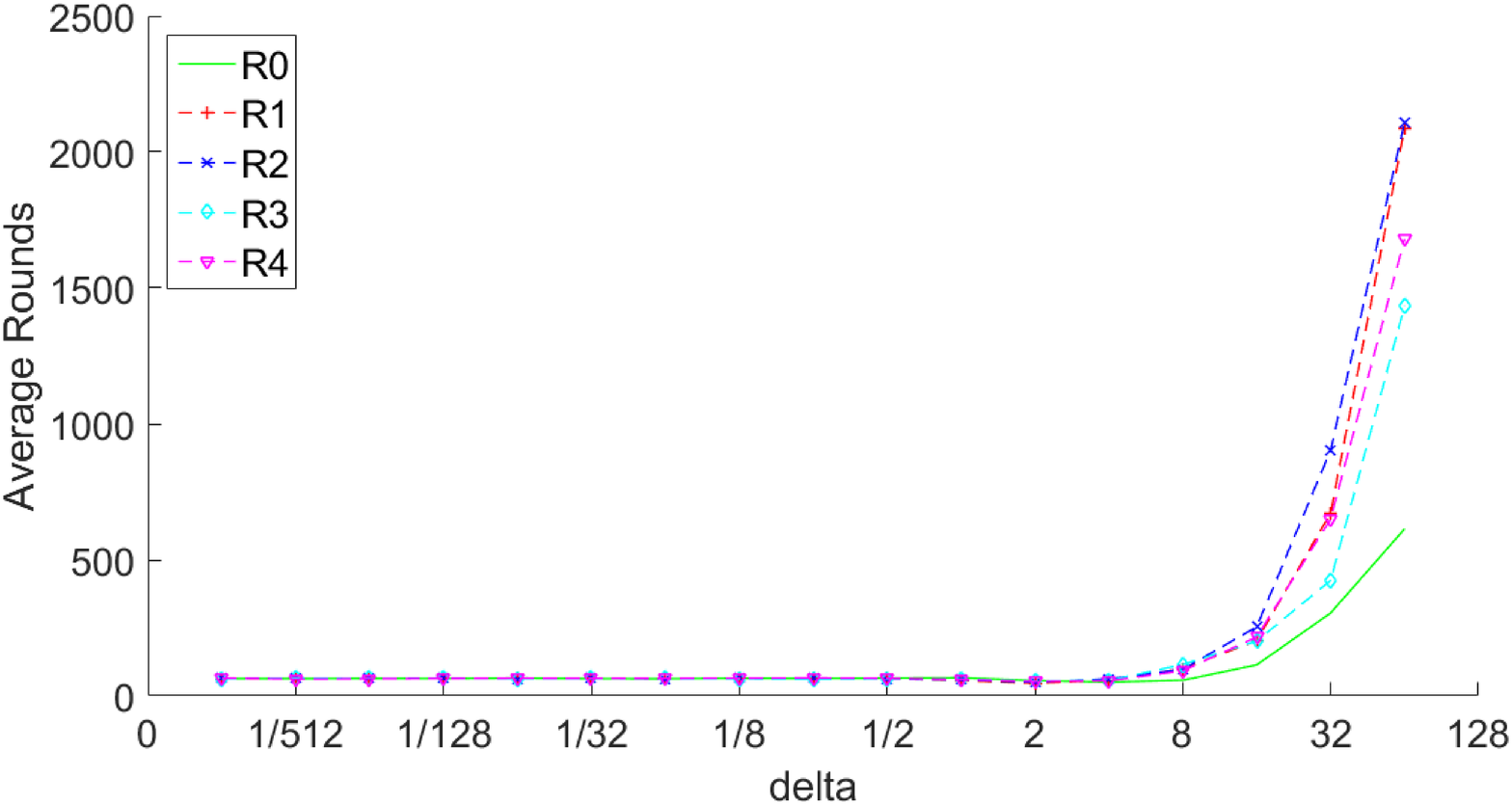}}
  \subfigure[Training set 3.]{\includegraphics[width=4cm]{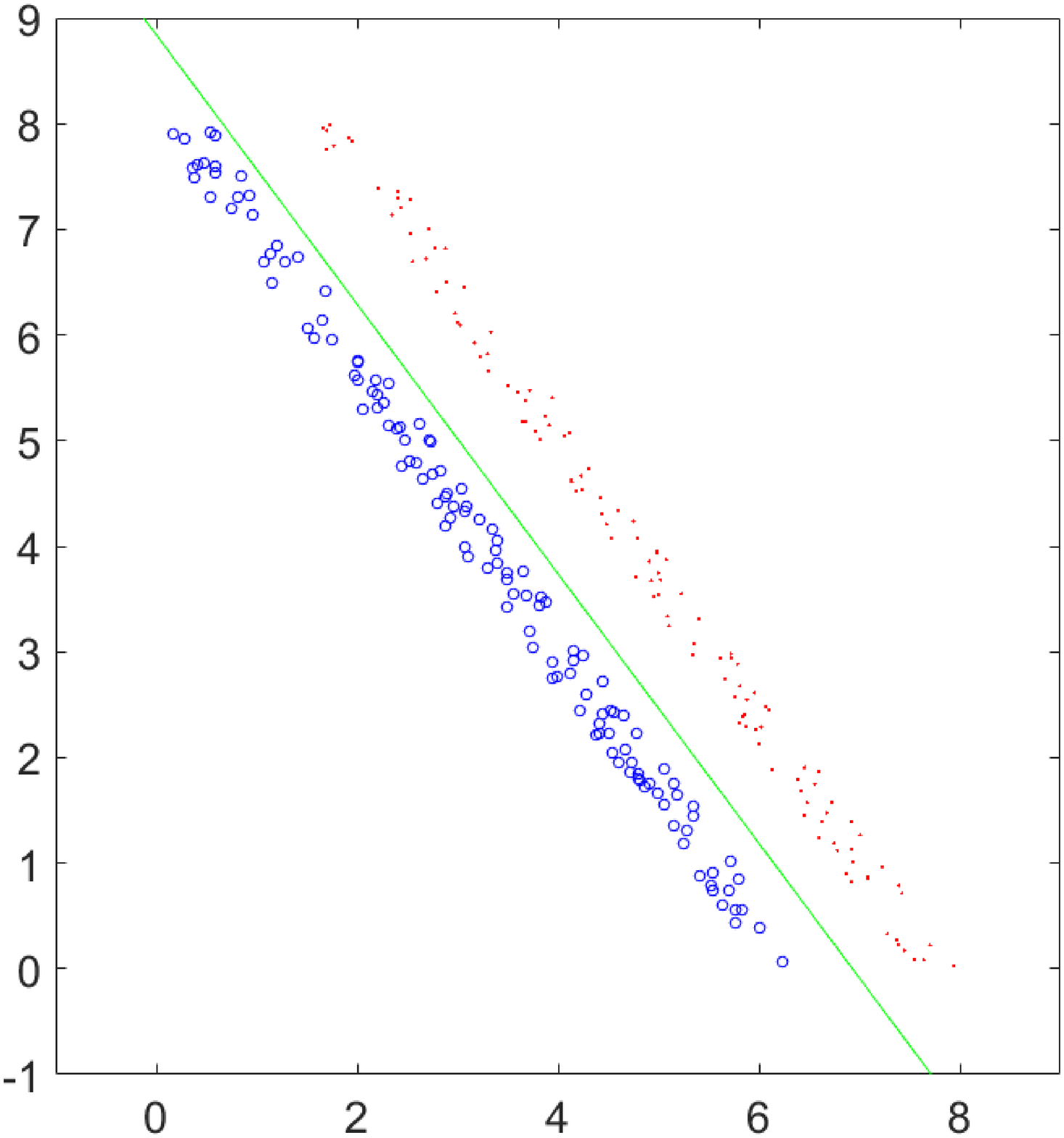}}
  \subfigure[Average running rounds on training set 3.]{\includegraphics[width=8cm]{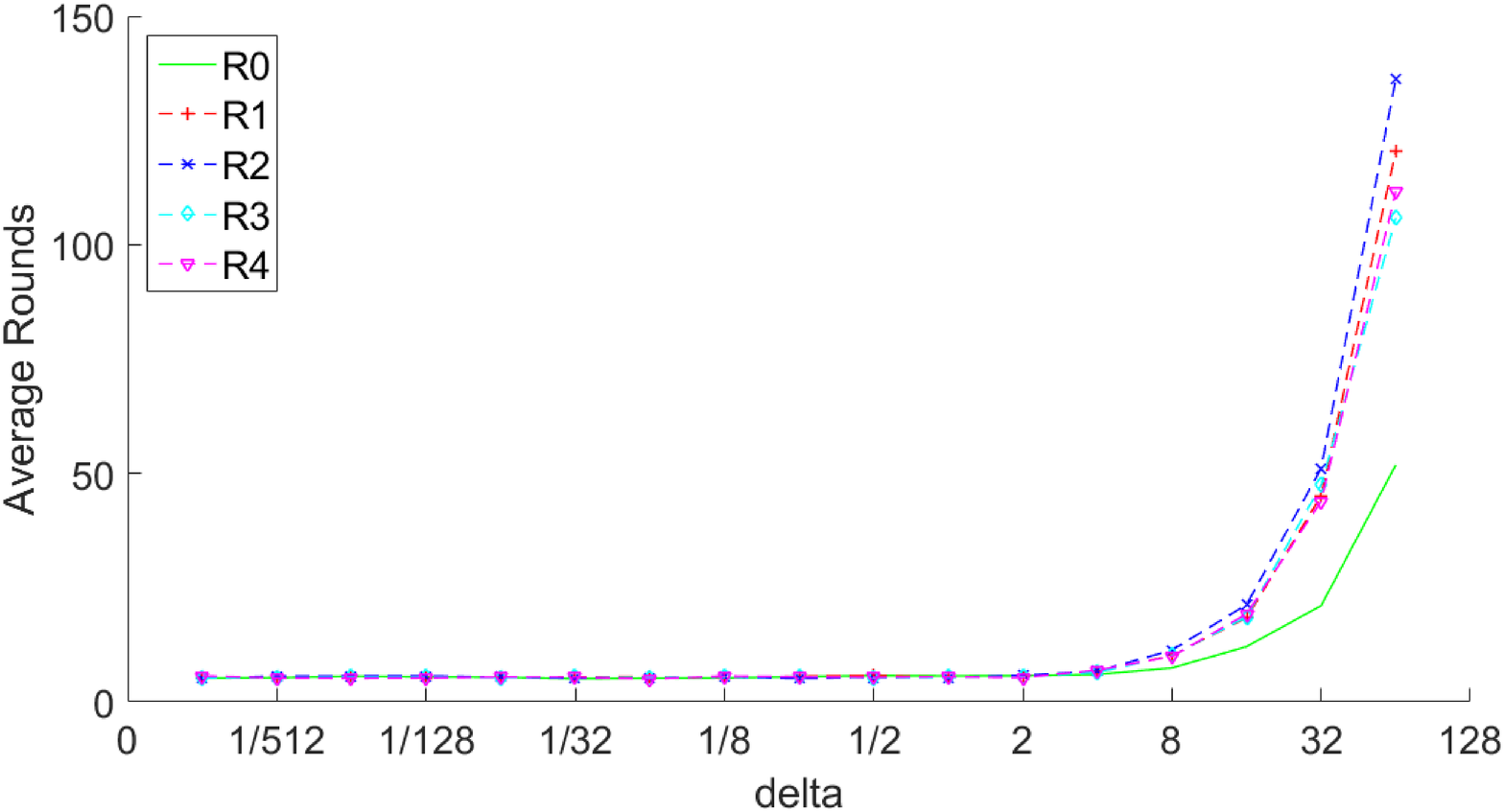}}
  \caption{(a, c, e): Training sets in Example \ref{Exam:1}. The (blue) circles and (red) dots are training examples corresponding to different sets respectively. The solid (green) lines are the classifier generated on this set by original perceptron algorithm (see Fig. \ref{Fig:Perceptron} or \cite{FreundS1999,Rosenblatt1957}). (b, d, f): Average running rounds of our quantum protocol on these training sets. $R_i$ means different random random number generator in Table \ref{Table:Generator}.}\label{Fig:Exam1}
\end{figure}

\begin{exam}\label{Exam:1}
    Consider three training sets shown in Fig. \ref{Fig:Exam1}(a), (c) and (e). The corresponding running rounds of Algorithm \ref{alg:ProtocolPerceptron} on these sets are given in subfigures (b), (d) and (f) respectively. The terminating and success probabilities for these situations are all 1. (See Section \ref{Sec:Comparison} for descriptions of terminating and success probabilities.)
    \begin{enumerate}
      \item Subfigure (a). In this case, training examples corresponding to different classes are independently generated by normal distributions on both coordinates.
      \item Subfigure (c). In this case, training examples are first generated by normal distributions on both coordinates. Then $w_0= (2,-1)$ and $b_0=-3$ are employed to set classes for all training examples.
      \item Subfigure (e). In this case, each $\vec{x}_i = (x_{i,1}, x_{i,2})$ and $c_i$ are generated in three steps: (1) uniformly at random choose  $r_1\in[0,0.5]$ and $r_2\in[0,8]$, (2) $x_{i,2}\la r_2$ and $x_{i,1}\la x_{i,2}+r_1$, (3) with a fifty-fifty chance, set $c_i=0$ or set $c_1=1$ and $x_{i,1}\la x_{i,1}+1.5$.
    \end{enumerate}
\end{exam}
In these figures, when the parameter $\delta$ is not big,  the running time (i.e., the average number of running rounds) may not change, and when $\delta$ is bigger, the running time increases with $\delta$. Therefore, the results show $\delta$ only influence the speed of our protocol. %Intuitively the greater $\delta$, the higher privacy level.

\subsection{Privacy Analysis for Quantum Protocol}\label{Sec:PrivacyAlice}
In this subsection, we present the privacy analysis of our quantum privacy-preserving perceptron protocol. We will first analyse the privacy metric of our quantum protocol, and then compare it with classical randomization methods.

\subsubsection{Quantum Protocol}
In our quantum protocol, there are mainly two ways to leak private information. The first one happens when computing $d=f(\vec{x}_{i\oplus u})$ at Step \textbf{S1}, and the second happens in the distorted training example $\vec{x}'$ used to update $w$ and $b$ at Step \textbf{S2}.

For Step \textbf{S1}, a quantitative analysis for each training example is given in Theorem \ref{thm:Probability}. The quantitative analysis for the entire database can be derived based on this. Recall in Theorem \ref{thm:Probability}, $n_1$ denotes the number of bits before the decimal mark of each attribute.
At this step, we say that Bob successfully attacks one training example at level $2^{n_1-n_2}$ with confidence 95\%, if he perform measurements with basis $\{|0\>,|1\>\}$ to reduce privacy amount of each attribute of this training example to $2^{n_1-n_2}$. Then from  Theorem \ref{thm:Probability}, we have the following:
\begin{cor}
    During the whole learning procedure in Fig. \ref{Fig:QPerceptron} (or Algorithm \ref{alg:ProtocolPerceptron}), the number of training examples that Bob can successfully attack at level $2^{n_1-n_2}$ with at least confidence 95\% before being detected is
    \begin{equation}\label{number}\frac{2n}{n_2}+\frac{2n}{n_2(n_2k-1)}-1\end{equation}
    on average.
\end{cor}
The number in (\ref{number}) is independent of the size $N$ of the training set. Usually $N\gg n$ for a large database. So only a tiny part of the entire database will be leaked at this step.

%For Step \textbf{S2}, since Bob does not know any information about the random number generator $R(\delta)$ and its parameter $\delta$, it is very hard for Bob to estimate the interval $[a,a+l]\ni x_{i,j}$. Theoretical analysis starts from the following fact:
%\begin{itemize}
%    \item Suppose Alice secretly holds $x\in [-L,L]$, a set $\mathcal{R}$ of random generators. And for any $R\in\mathcal{R}$, any random number $R$ generated by $R$ is in $[-L,L]$.
%     Then Alice sends $x'=x+r$ to Bob. If $x'$ is still in $[-L,L]$, Bob cannot estimate $x\in[a,a+l]$ with confidence 95\% and $l<1.9L$.
%\end{itemize}
%This is because for any $x'\in [-L,L]$ and $x\in[-L,L]$, there exists many random generators generating same random number $x'-x$. Therefore if the range of some attribute is $[a,b]$, Alice can publish a range $[-L,L]\supseteq [a,b]$ and employs a secret random number generator to protect her privacy.

Now we consider Step \textbf{S2}. The most important thing to protect privacy in this step is  that
\begin{itemize}
  \item Alice keeps her random number generator $R(\delta)$ and the parameter $\delta$ secret.
\end{itemize}
This fact ensures that Bob is not able to get the distribution of $x$ directly from $x'$. We can explain it by a simple example. Suppose original data $X$ is a uniformly distributed random variable, and noise $Y$ is a random variable with a normal distribution. Alice publishes some samples of distorted data $X+Y$. Then since Bob does not know any information about $Y$, he cannot distinguish the following two situations:
\begin{enumerate}
  \item The original data is $X$, and the noise is $Y$.
  \item The original data is $Y$, and the noise is $X$.
\end{enumerate}
Obviously, there are many other variables $X'$ and $Y'$ with various distributions, whose composition $X'+Y'$ has the same distribution with $X+Y$. Therefore, Bob cannot get any information about $\vec{x}_i$ simply from $\vec{x}'_i$, if Alice keeps her random number generator secret.

For convenience of the comparison in the following subsection, we need to quantify the privacy amount in Step \textbf{S2}. Note that if Bob does not cheat, the privacy amount for $\vec{x}_i$ is at least $0.95\delta$ with confidence 95\%. This is because (1) in our protocol, Bob does not know any information about random number generator $R(\delta)$, and (2) even if he knows $\delta$ and $R(\delta)$ is $R_0(\delta)$, the privacy amount is $1.9\delta/2=0.95\delta$ with confidence 95\% ($R_0(\delta)$ leads to $1.9\delta$, and one-bit information of $f(\vec{x}_i)$ halves it).

\subsubsection{Comparison with the Classical Methods}\label{Sec:Comparison}
In this subsection, we compare our quantum protocol with the known classical methods for a similar task. In the two-party situation, a classical method, for instance \cite{AgrawalS2000}, works in the following way:
\begin{enumerate}
  \item \textbf{Randomization}. Alice uses random number generator $R$ to add noise into $D$ to produce a distorted database $D'$. Then Alice publishes $R$ and $D'$ to Bob.
  \item \textbf{Reconstruction}. Bob employs $R$ and $D'$ to reconstruct an estimated distribution $\tilde{D}$ of original training examples in $D$.
  \item \textbf{Learning}. Bob finishes the learning task on $\tilde{D}$.
\end{enumerate}
In \cite{AgrawalS2000}, it is shown that the reconstruction method works well for some situations. Unfortunately, it does not work for our problem, for instance, training sets in Fig. \ref{Fig:Exam1}(c) and \ref{Fig:Exam1}(e). This is because the reconstruction in \cite{AgrawalS2000} requires that
for every class, different attributes are not strongly related.
%It is best that every attribute is independent to others.
Otherwise, the one-dimensional reconstruction method in \cite{AgrawalS2000} will not correctly estimate the distributions of attributes one by one. Since it is beyond the scope of this paper, we are not going to further discuss this weakness.

An alternative method follows \cite{AgrawalS2000}. But one step further, distributions of all attributes are estimated once together, not one by one. This method works much better, but still worse than our quantum protocol. Moreover, the most important weakness of this method is its complexity. It will be \textit{exponential} on the number $k$ of attributes. See Appendix \ref{Apd:Reconstruction} for more discussions of these reconstruction methods.

\begin{figure}
  \centering
  \subfigure[Terminating probabilities of training set 1.]{\includegraphics[width=6cm]{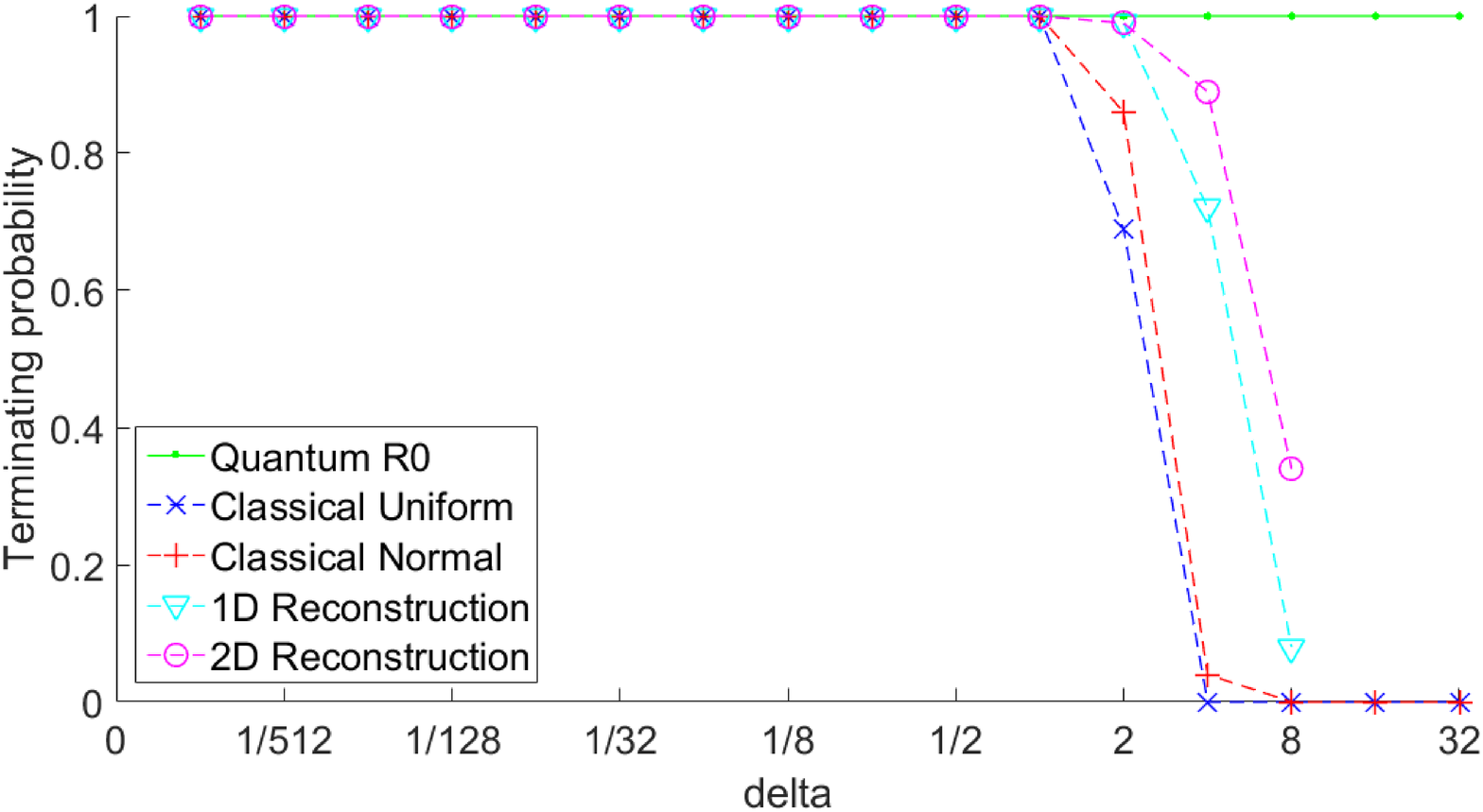}}
  \subfigure[Success probabilities of training set 1.]{\includegraphics[width=6cm]{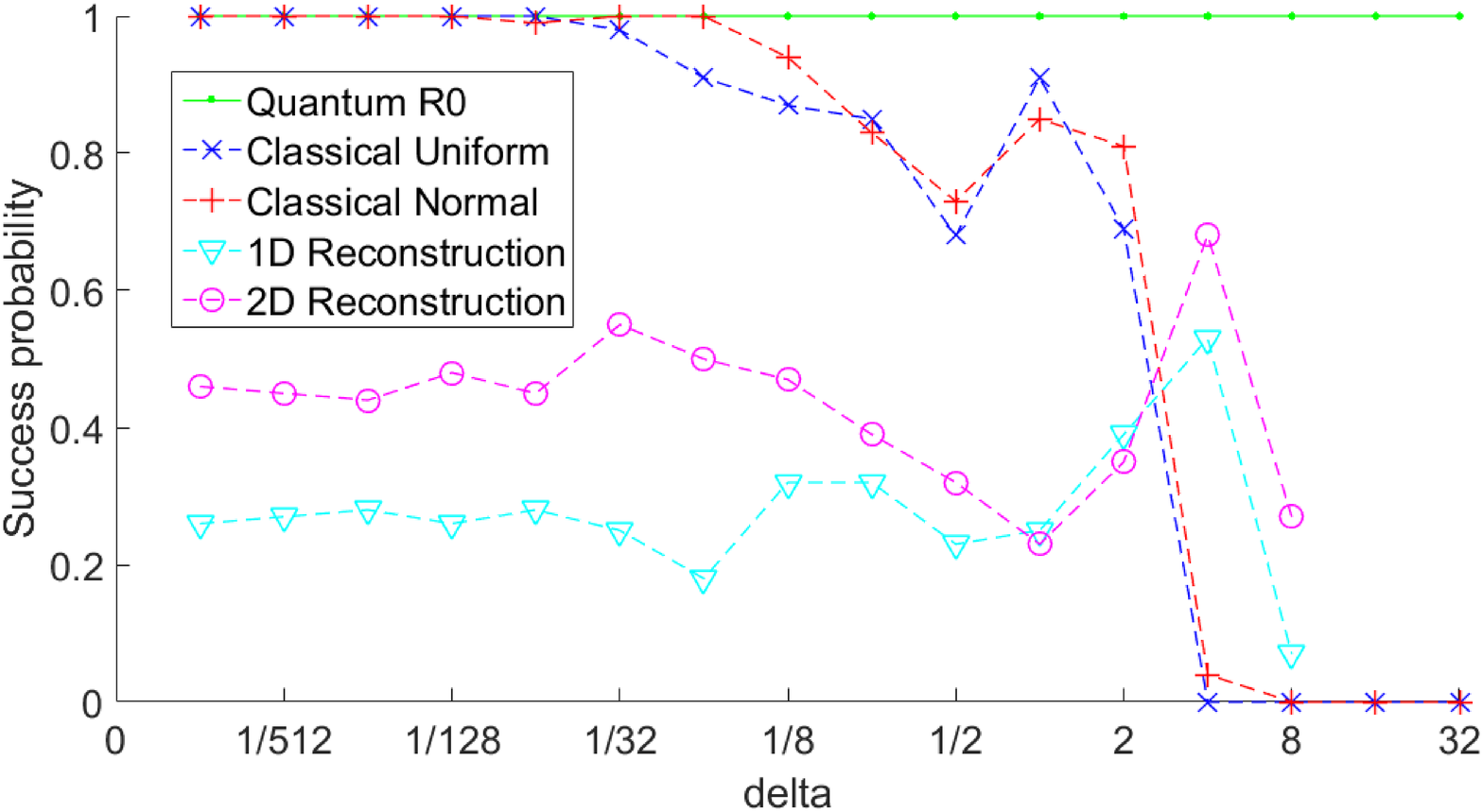}}
  \subfigure[Terminating probabilities of training set 2.]{\includegraphics[width=6cm]{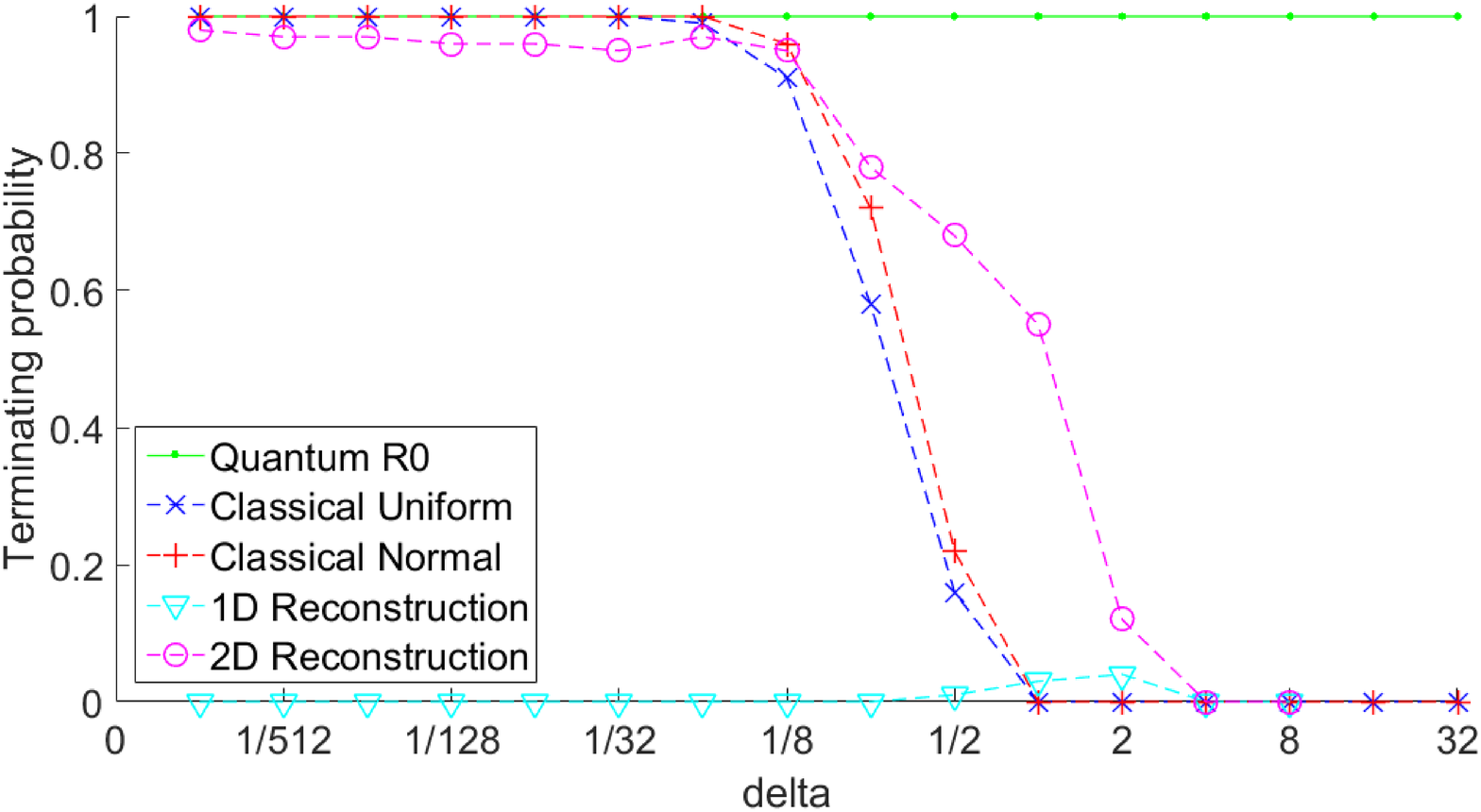}}
  \subfigure[Success probabilities of training set 2.]{\includegraphics[width=6cm]{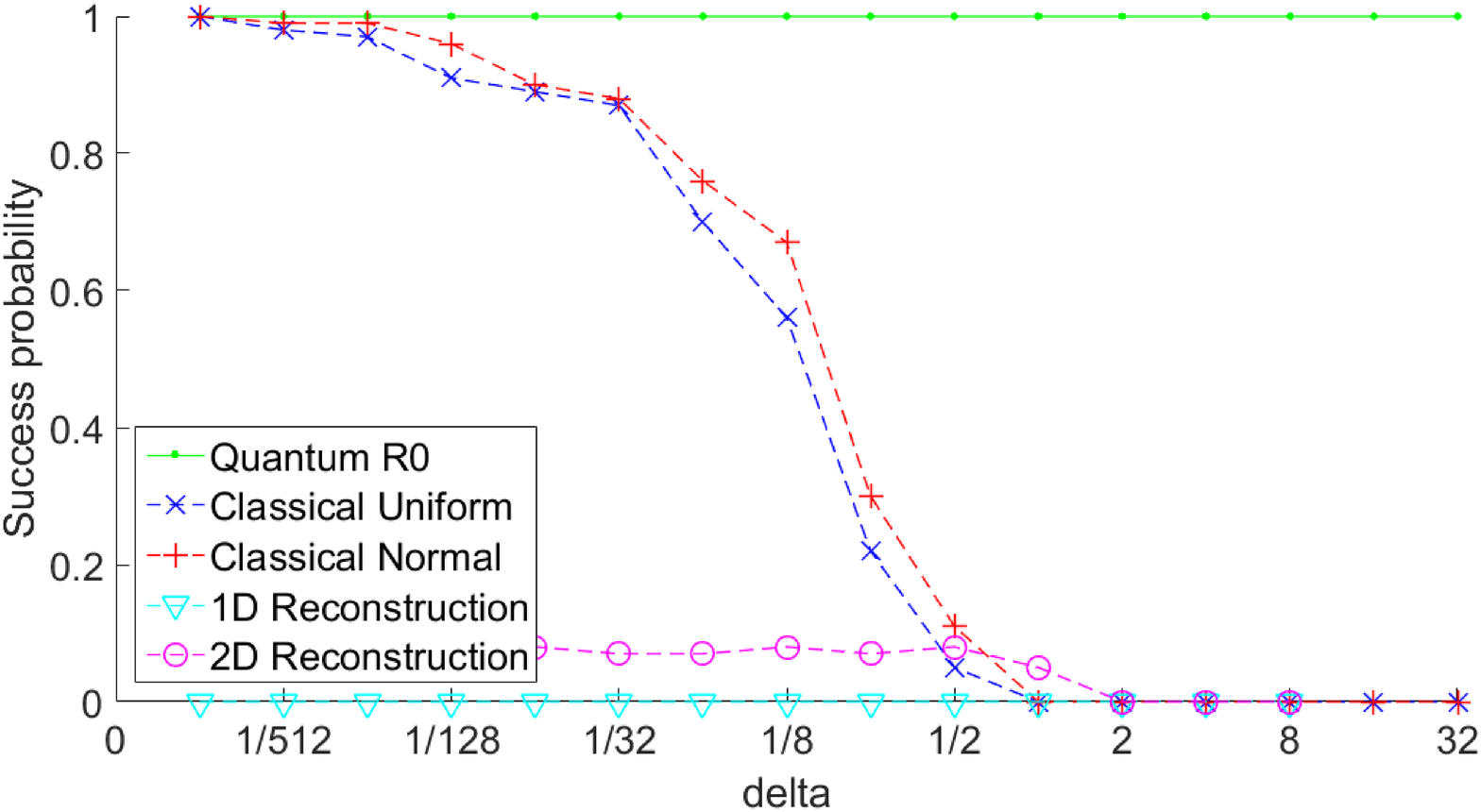}}
  \subfigure[Terminating probabilities of training set 3.]{\includegraphics[width=6cm]{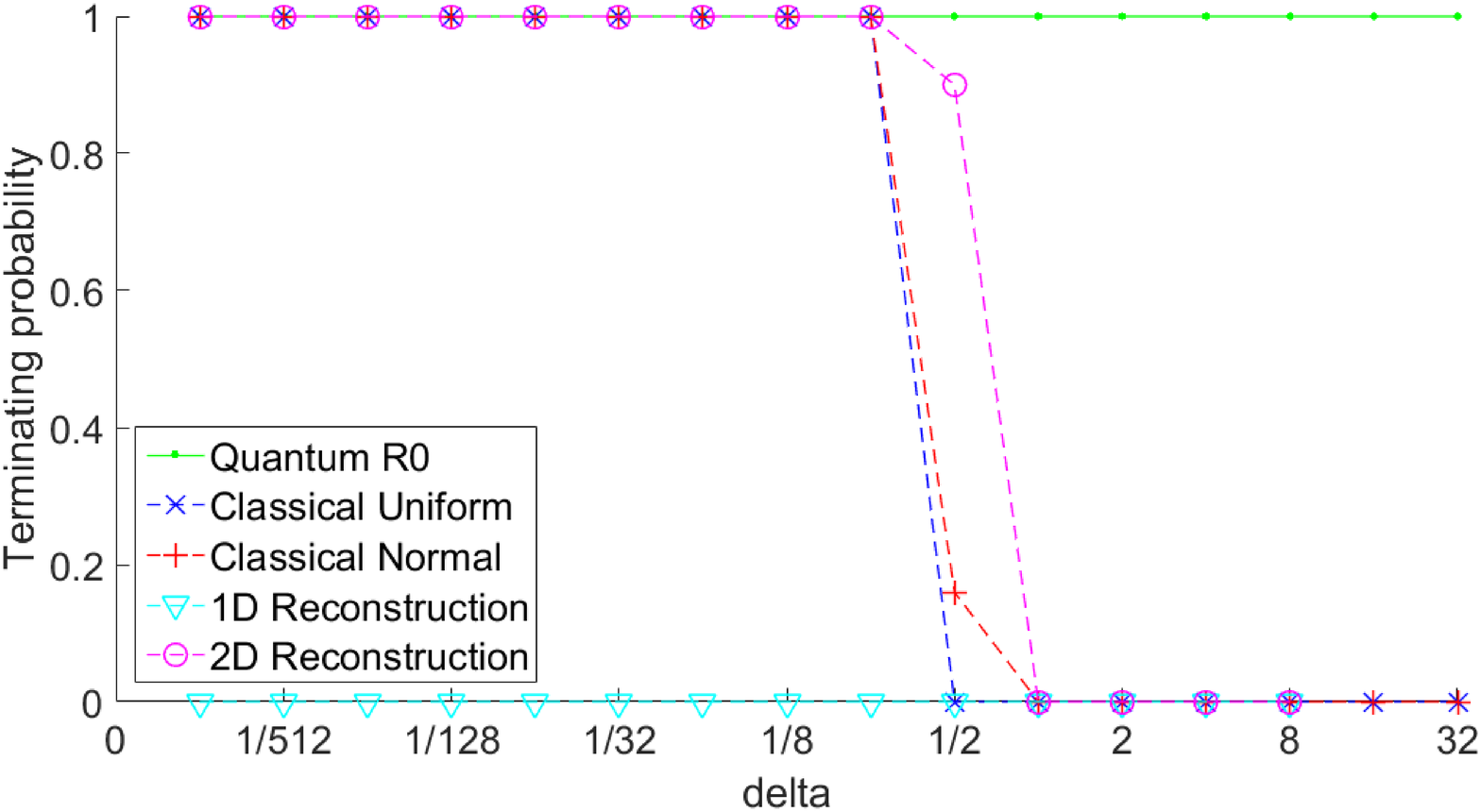}}
  \subfigure[Success probabilities of training set 3.]{\includegraphics[width=6cm]{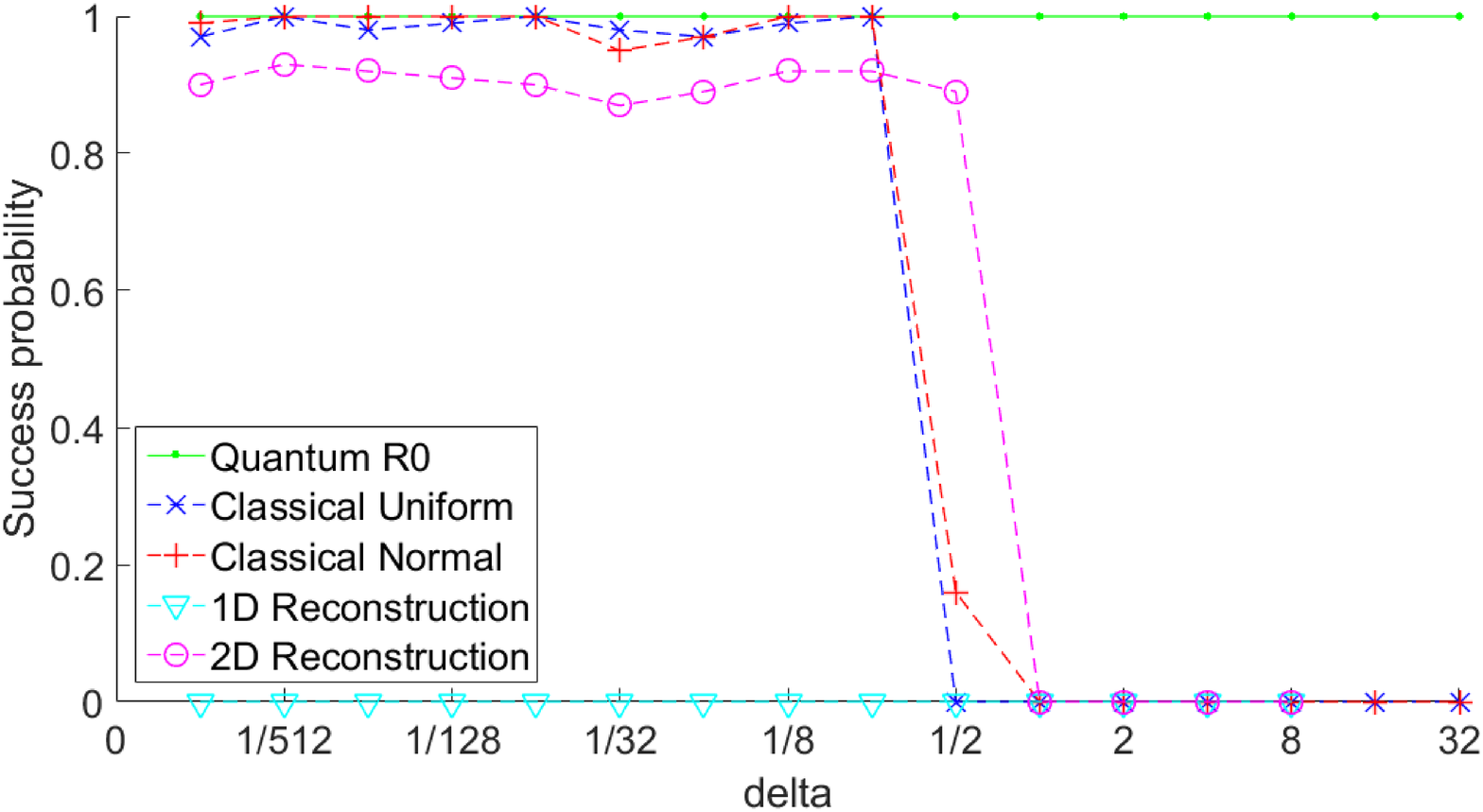}}
  \caption{The comparison of our quantum protocol with four classical methods on the training sets in Example \ref{Exam:1}. All classical methods with parameter $\delta$ here have the same privacy level $1.9\delta$ with confidence 95\%. Our quantum protocol i.e., $R_0$ in the figures, has privacy level at least $1.9\delta$ with confidence 95\%.}\label{Fig:Exam1Com}
\end{figure}

Due to the weakness of the reconstruction, let us compare our quantum protocol with classical randomization methods without reconstruction. Totally, we consider the following four methods:
\begin{itemize}
  \item Uniformly distributed noise from $[-\delta,\delta]$ without reconstruction.
  \item Normally distributed noise of mean 0 and standard deviation $0.484\delta$ without reconstruction.
  \item Uniformly distributed noise from $[-\delta,\delta]$ with reconstruction for attributes one by one.
  \item Uniformly distributed noise from $[-\delta,\delta]$ with reconstruction for attributes once together.
\end{itemize}
As stated in \cite{AgrawalS2000}, all these methods has the same privacy amount $1.9\delta$ with confidence 95\%. The comparison is based on the terminating probabilities and success probabilities:
\begin{itemize}
  \item Termination. We say that one execution of a classical or quantum protocol terminates, if and only if it terminates in $T=40000$ outer loops.
  \item Success. We say that one execution  of a classical or quantum protocol succeeds, if and only if it terminates (in 40000 outer loops) and the resulting classifier classifies $D$ correctly.
\end{itemize}
The comparison is given in Fig. \ref{Fig:Exam1Com}. The results show that when we require high privacy level, our quantum protocol still works perfectly, but classical methods can hardly work. This indicates that our quantum protocol can preserve privacy much better than classical methods.

\section{Conclusion and Discussion}\label{Sec:Conclusion}
In this paper, we develop a privacy-preserving quantum algorithm to compute a general linear function $f(\vec{x})$ and then apply it to machine learning based on perceptron. The performance of our algorithm has been analysed both theoretically and numerically. The main features of our algorithm include:\begin{itemize}
\item
\textbf{Classical databases versus quantum database}: The quantum algorithms for search \cite{Grover1996}, counting \cite{BrassardHT1998}, machine learning \cite{KapoorWS2016Perceptron,LloydMR2013,RebentrostML2014,WiebeKS2014DL} and private queries \cite{GiovannettiLM2008PQ} all require a quantum database, for instance, in the form of quantum random access memory \cite{GiovannettiLM2008qram}. However, our quantum algorithm is executed on classical databases, and no quantum database is required.  So, our algorithm is much easier to implement.
\item \textbf{Less entanglement}: For example, the quantum algorithms presented in \cite{Grover1996,BrassardHT1998,KapoorWS2016Perceptron} require that all the working qubits are entangled, and it lasts for $O(\sqrt{N})$ rounds. In contrast, our algorithm requires much less entangled qubits, and the entanglement during the communication and computation only needs to last very shortly for one round. Indeed, in calling procedure \ref{alg:ProcedureCompute}  to compute $f(\vec{x})$ in Algorithm \ref{alg:ProtocolComputeF}, Alice sends a separable state $|+\>|\vec{x}\>$ to Bob and Bob sends $|+\>|\vec{x}\>$ or $|-\>|\vec{x}\>$ back to Alice. There is no entanglement during this communication. Meanwhile, when Bob computes $f(\vec{x})$ on this separable state, Bob may reduce the number of entangled qubits to a small number, say $n$. Moreover, this entanglement only needs to last for one computational round. Even in a test which employs certain entangled states to detect dishonesty, only two qubits are entangled during the communication, and the entanglement only lasts for one round as well. So, our algorithm might be implementable using the intermediate quantum computing technology.\end{itemize}

To conclude the paper, we consider some possibilities of extending our algorithm for more applications. First, it can be applied to a more general artificial neural network; for example, a feed forward neural network employing nonlinear neurons \cite{Nielsen2015}. Since these nonlinear functions can be decomposed into a linear function followed by a nonlinear one, our algorithm can be directly used to the linear part of the input level of a network. A similar idea applies to the backpropagation algorithm \cite{Nielsen2015}. Second, our algorithm can be generalised to deal with several machine learning and data mining tasks where a linear or boolean function is used, for instance decision tree learning \cite{AgrawalS2000,Quinlan1986} and association rule mining \cite{EvfimievskiSAG2004}.

{%\small
\bibliography{RefQP01}

\providecommand{\noopsort}[1]{}\providecommand{\singleletter}[1]{#1}%
\begin{thebibliography}{10}

\bibitem{AgrawalES2003}
Rakesh Agrawal, Alexandre Evfimievski, and Ramakrishnan Srikant.
\newblock Information sharing across private databases.
\newblock In {\em Proceedings of the 2003 ACM SIGMOD International Conference
  on Management of Data}, SIGMOD '03, pages 86--97, New York, NY, USA, 2003.
  ACM.

\bibitem{AgrawalS2000}
Rakesh Agrawal and Ramakrishnan Srikant.
\newblock Privacy-preserving data mining.
\newblock In {\em Proceedings of the 2000 ACM SIGMOD International Conference
  on Management of Data}, SIGMOD '00, pages 439--450, New York, NY, USA, 2000.
  ACM.

\bibitem{AharonovTVY2000}
Dorit Aharonov, Amnon Ta-Shma, Umesh~V Vazirani, and Andrew~C Yao.
\newblock Quantum bit escrow.
\newblock In {\em Proceedings of the thirty-second annual ACM symposium on
  Theory of computing}, pages 705--714. ACM, 2000.

\bibitem{BennettB1984}
Charles~H Bennett and Gilles Brassard.
\newblock Quantum cryptography: Public key distribution and coin tossing.
\newblock {\em Theoretical computer science}, 560:7--11, 2014.

\bibitem{BrassardHT1998}
Gilles Brassard, Peter H{\o}yer, and Alain Tapp.
\newblock Quantum counting.
\newblock In KimG. Larsen, Sven Skyum, and Glynn Winskel, editors, {\em
  Automata, Languages and Programming}, volume 1443 of {\em Lecture Notes in
  Computer Science}, pages 820--831. Springer Berlin Heidelberg, 1998.

\bibitem{Chaudhuri2009}
Kamalika Chaudhuri and Claire Monteleoni.
\newblock Privacy-preserving logistic regression.
\newblock In {\em Advances in Neural Information Processing Systems}, pages
  289--296, 2009.

\bibitem{ColbeckK2006}
Roger Colbeck and Adrian Kent.
\newblock Variable-bias coin tossing.
\newblock {\em Phys. Rev. A}, 73:032320, Mar 2006.

\bibitem{EvfimievskiSAG2004}
Alexandre Evfimievski, Ramakrishnan Srikant, Rakesh Agrawal, and Johannes
  Gehrke.
\newblock Privacy preserving mining of association rules.
\newblock {\em Information Systems}, 29(4):343 -- 364, 2004.

\bibitem{FreundS1999}
Yoav Freund and Robert~E Schapire.
\newblock Large margin classification using the perceptron algorithm.
\newblock {\em Machine learning}, 37(3):277--296, 1999.

\bibitem{GiovannettiLM2008PQ}
Vittorio Giovannetti, Seth Lloyd, and Lorenzo Maccone.
\newblock Quantum private queries.
\newblock {\em Phys. Rev. Lett.}, 100:230502, Jun 2008.

\bibitem{GiovannettiLM2008qram}
Vittorio Giovannetti, Seth Lloyd, and Lorenzo Maccone.
\newblock Quantum random access memory.
\newblock {\em Phys. Rev. Lett.}, 100:160501, Apr 2008.

\bibitem{Grover1996}
Lov~K. Grover.
\newblock A fast quantum mechanical algorithm for database search.
\newblock In {\em Proceedings of the Twenty-eighth Annual ACM Symposium on
  Theory of Computing}, STOC '96, pages 212--219, New York, NY, USA, 1996. ACM.

\bibitem{HardyK2004}
Lucien Hardy and Adrian Kent.
\newblock Cheat sensitive quantum bit commitment.
\newblock {\em Phys. Rev. Lett.}, 92:157901, Apr 2004.

\bibitem{KapoorWS2016Perceptron}
Ashish Kapoor, Nathan Wiebe, and Krysta Svore.
\newblock Quantum perceptron models.
\newblock In {\em Advances in Neural Information Processing Systems}, pages
  3999--4007, 2016.

\bibitem{LloydMR2013}
Seth Lloyd, Masoud Mohseni, and Patrick Rebentrost.
\newblock Quantum algorithms for supervised and unsupervised machine learning.
\newblock {\em arXiv preprint arXiv:1307.0411}, 2013.

\bibitem{Nielsen2015}
Michael~A Nielsen.
\newblock Neural networks and deep learning.
\newblock {\em URL: http://neuralnetworksanddeeplearning.com/}, 2015.

\bibitem{NielsenC2010}
Michael~A Nielsen and Isaac Chuang.
\newblock {\em Quantum computation and quantum information}.
\newblock Cambridge, 2010.

\bibitem{Novikoff1963}
Albert~BJ Novikoff.
\newblock On convergence proofs for perceptrons.
\newblock Technical report, DTIC Document, 1963.

\bibitem{Quinlan1986}
J.~Ross Quinlan.
\newblock Induction of decision trees.
\newblock {\em Machine learning}, 1(1):81--106, 1986.

\bibitem{RebentrostML2014}
Patrick Rebentrost, Masoud Mohseni, and Seth Lloyd.
\newblock Quantum support vector machine for big data classification.
\newblock {\em Phys. Rev. Lett.}, 113:130503, Sep 2014.

\bibitem{Ricci2011}
Francesco Ricci, Lior Rokach, and Bracha Shapira.
\newblock {\em Introduction to recommender systems handbook}.
\newblock Springer, 2011.

\bibitem{Rosenblatt1957}
Frank Rosenblatt.
\newblock The perceptron: A probabilistic model for information storage and
  organization in the brain.
\newblock {\em Psychological review}, 65(6):386, 1958.

\bibitem{VaidyaC2002}
Jaideep Vaidya and Chris Clifton.
\newblock Privacy preserving association rule mining in vertically partitioned
  data.
\newblock In {\em Proceedings of the eighth ACM SIGKDD international conference
  on Knowledge discovery and data mining}, pages 639--644. ACM, 2002.

\bibitem{VanderVaart2000asymptotic}
Aad~W Van~der Vaart.
\newblock {\em Asymptotic statistics}.
\newblock Cambridge university press, 2000.

\bibitem{Wichura2001}
Michael~J. Wichura.
\newblock Lecture notes.
\newblock {\em https://galton.uchicago.edu/\textasciitilde
  wichura/Stat304/Handouts/L12.cf2.pdf}.

\bibitem{WiebeKS2014DL}
Nathan Wiebe, Ashish Kapoor, and Krysta~M Svore.
\newblock Quantum deep learning.
\newblock {\em arXiv preprint arXiv:1412.3489}, 2014.

\bibitem{Williams1991probability}
David Williams.
\newblock {\em Probability with martingales}.
\newblock Cambridge university press, 1991.

\bibitem{YingYF2015}
Shenggang Ying, Mingsheng Ying, and Yuan Feng.
\newblock Quantum privacy-preserving data mining.
\newblock {\em arXiv preprint arXiv:1512.04009v2}, 2015.

\bibitem{YingYF2017}
Shenggang Ying, Mingsheng Ying, and Yuan Feng.
\newblock Quantum privacy-preserving data analytics.
\newblock {\em arXiv preprint arXiv:1702.04420}, 2017.

\end{thebibliography}
}

\newpage
\appendix

\section{Proofs for Theorems}

\subsection{Proof of Theorem \ref{thm:SecurityA1}}
The proof depends on three lemmas. The proofs of these lemmas are organised in several subsections. We starts from the following lemma.
\begin{lem}\label{lem:State}
    Suppose Alice sends $|\psi(\vec{y},u,m)\>$ to Bob. Here $u$ and $m$ may be 0 to represent a computational state. If Bob wants to always pass tests with certainty, the states he sends back to Alice should be separable from any other additional quibts. Moreover it should be either $|\psi(\vec{y},u,m)\>$ or $Z_0|\psi(\vec{y},u,m)\>$, where $Z_0$ is the $Z$-gate on the result qubit.
\end{lem}
From this lemma, one can find that if Bob wants to always pass tests with certainty by directly sending back the states after his actions, the only operator that he can perform is an identity operator or a $Z$-gate on the result qubit. This indicates that
\begin{itemize}
  \item Bob cannot get private information and always pass tests with certainty by directly sending back states simultaneously.
\end{itemize}

What happens if he performs measurements to read private information, and then constructs a new state and sends it back to Alice? We can prove that he cannot always pass tests in this way. The proof is based on the following two lemmas.
\begin{lem}\label{thm:TestStateDis}
    Suppose Bob constructs a measurement $\{M_v:v\}$ to distinguish the test states. If Bob gets measurement outcome $v$, then the probability that $|\psi(y,u,m)\>$ is the test state is at most $\frac{1}{nk}$, i.e.
    \begin{equation*}
        \Pr(|\psi(\vec{y},u,m)\>~|~M_v)\leq \frac{1}{nk}.
    \end{equation*}
\end{lem}
\begin{lem}\label{thm:SuccProb}
    Suppose Bob gets measurement outcome $v$ and constructs a test state based on this outcome. If he sends back this new state back to Alice, the success probability to pass the test is at most $\frac{nk+3}{4nk}$.
\end{lem}
Therefore, we conclude that
\begin{itemize}
  \item Bob cannot pass tests with certainty by performing measurements and then reconstructing a test state.
\end{itemize}
and thus complete the proof.

\subsubsection{Proof of Lemma \ref{lem:State}}\label{Sec:LemState}
We first prove that the $nk+1$ qubits that Bob sends back to Alice should be separable from any other additional qubits, if Bob wants to always pass tests with certainty. Since any mixed state can be purified by adding a new system, we assume without loss of generality Bob holds state
\begin{equation*}
    |\phi\>=\sum \alpha_{i,z}|i\>|z\>|\xi_{i,z}\>
\end{equation*}
after his actions, where $i\in\{0,1\}$ and $z\in\{0,1\}^{nk}$. Then Bob sends the first $nk+1$ bits (i.e., the first two subsystems) back to Alice. After $U_{CNOT} Z_0^u U_{SWAP}(1,m)$, Alice gets the first $nk+1$ qubits of $U_{CNOT} Z_0^u U_{SWAP}(1,m)\otimes I|\phi\>$. Since Line \ref{line:CM} in procedure \ref{alg:ProcedureCompute} should be always passed, the state on the data qubits should be $|\vec{y}\>$ with certainty. So this state has the following form:
\begin{equation*}
    U_{CNOT} Z_0^u U_{SWAP}(1,m)\otimes I |\phi\>=\beta|+\>|\vec{y}\>|\xi_0\>+\tau|-\>|\vec{y}\>|\xi_1\>.
\end{equation*}
Moreover, in order to pass Line \ref{line:TM} in Algorithm \ref{alg:ProtocolComputeF}, we should have $|\beta|=1$ or $|\tau|=1$. As the operator $U_{CNOT} Z_0^u U_{SWAP}(1,m)$ only works on the result and data qubits, $|\phi\>$ can be rewritten as
\begin{equation}\label{Eq:LemmaA1E1}
    |\phi\>=\sum \alpha_{i,z}|i\>|z\>|\xi\>
    = \begin{cases} (U_{SWAP}(1,m)Z_0^u U_{CNOT}|+\>|\vec{y}\>)|\xi\>, \\ (U_{SWAP}(1,m)Z_0^u U_{CNOT}|-\>|\vec{y}\>)|\xi\>. \end{cases}
\end{equation}
Thus, the first $nk+1$ qubits are isolated.

Secondly we prove that Bob's operators are equivalent to $I$ or $Z_0$. Observe that the operator $Z_0$ commutes with $U_{SWAP}(1,m)Z_0^u U_{CNOT}$ and $|\psi(\vec{y},u,m)\>= (U_{SWAP}(1,m)Z_0^u U_{CNOT} )|+\>|\vec{y}\>$. So, from Eq. \eqref{Eq:LemmaA1E1} we have
\begin{equation*}
    |\phi\> =  |\psi(\vec{y},u,m)\>|\xi\>, \text{~or~} |\phi\> = Z_0|\psi(\vec{y},u,m)\>|\xi\>.
\end{equation*}
This completes the proof.

\subsubsection{Proof of Lemma \ref{thm:TestStateDis}}

 The idea of this proof comes from the fact that $B_{m'}=\{|\psi(\vec{y}',u',m')\>:\forall \vec{y}',u'\}$ forms an orthonormal basis for each $m'$.

    In this proof, we use $q=\Pr(|\psi(\vec{y},u,m)\>)$ to denote the probability that the test state is $|\psi(\vec{y},u,m)\>$. Since the test is generated uniformly at random, we have: $$q=\Pr(|\psi(\vec{y},u,m)\>)=\frac{1}{2nk2^{nk}}$$ for any test state.
On the other hand, the probability $\Pr(M_v~|~|\psi(\vec{y},u,m)\>)$ that the outcome is $v$ when the current state is $|\psi(\vec{y},u,m)\>$ is
    \begin{equation*}
        \Pr(M_v~|~|\psi(\vec{y},u,m)\>)=\tr(M_v^\dag M_v|\psi(\vec{y},u,m)\>\<\psi(\vec{y},u,m)| ).
    \end{equation*}
    Consequently, the probability $\Pr(M_v)$ that the outcome is $v$ is
    \begin{align*}
        \Pr(M_v)&=\sum_{\vec{y}',u',m'} \Pr(M_v~|~|\psi(\vec{y}',u',m)\>)\Pr(|\psi(\vec{y}',u',m')\>)\\
        &=\sum_{\vec{y}',u',m'} q\tr(M_v^\dag M_v|\psi(\vec{y}',u',m')\>\<\psi(\vec{y}',u',m')| )\\
        &=\sum_{m'} q\tr(M_v^\dag M_v )=nkq\tr(M_v^\dag M_v).
    \end{align*}
    In the above equation, the last step comes from the fact that $B_{m'}=\{|\psi(\vec{y}',u',m')\>:\forall \vec{y}',u'\}$ is an orthonormal basis.
    Therefore, we have:
    \begin{align*}
        \Pr(|\psi(\vec{y},u,m)\>~|~M_v)&\leq \sum_{\vec{y}',u'} \Pr(|\psi(\vec{y}',u',m)\>~|~M_v)\\
        &=\sum_{\vec{y}',u'}\frac{\Pr(M_v~|~|\psi(\vec{y}',u',m)\>)\Pr(|\psi(\vec{y}',u',m)\>)}{\Pr(M_v)}\\
        &=\frac{1}{nk}.
    \end{align*}
    This completes the proof.

\subsubsection{Proof of Lemma \ref{thm:SuccProb}}

    Suppose the correct state is $|\psi(\vec{y},u,m)\>$, and Bob constructs $|\psi(\vec{y}',u',m')\>$. We consider the following two cases:

        \smallskip\

    \textbf{Case 1}. $m=m'$: In this case, Bob can construct a correct test state with $\vec{y}=\vec{y}'$ and $u=u'$, as he can perform measurements to read out $\vec{y}'$ and $u'$. The success probability is 1.

    \smallskip\

    \textbf{Case 2}. $m\neq m'$: Without loss of generality, we assume $\vec{y}=\vec{0}$, $u=0$, $m=1$ and $m'=2$. Then with probability 0.5, Alice gets measurement outcomes $0$ on the result qubit and $\vec{0}$ on the data qubits. Or with probability 0.5, Alice gets measurement outcomes $1$ on the result qubit and $1110\cdots$ on the data qubits. The latter situation will be detected at Line \ref{line:CM}. %Moreover, for the other test states, this situation happens again.
    So the total probability to pass the test is $0.5*0.5=0.25$.

    \smallskip\

    By Theorem \ref{thm:TestStateDis} and its proof, the first case happens with probability at most $\frac{1}{nk}$. So in conclusion, the success probability in total is at most
    $$\frac{1}{nk}*1+\frac{nk-1}{nk}*\frac{1}{4}=\frac{nk+3}{4nk}.$$

\subsection{Proof of Theorem \ref{thm:Probability}}
Firstly, since Bob wants to estimate $\vec{x}_i$ or $\vec{x}_{i,j}$ with confidence at least 95\%, he has to perform measurements in each Procedure \ref{alg:ProcedureCompute}. Otherwise, he has at least $\frac{1}{3}$ probability to miss the true input data $\vec{x}_i$. So Bob performs measurements on two test states $|\psi(\vec{y},u,m)\>$.

Secondly, it is assumed the measurement basis is $\{|0\>,|1\>\}$ because of the 95\% confidence. Once Bob performs measurements on a qubit of a state, he gets one-bit information about the state, and thus halves the privacy amount of this state. So besides the one-bit information from the result $f(\vec{x}_i)$, Bob still needs $n_2-1$ bits for a single attribute (respectively  $n_2 k-1$ bits for an entire training example).

Thirdly, the probability that Bob's cheat is detected is decided by the probability that measurements are performed on the chosen $m$-th qubit in test states. Since $m$ is chosen uniformly at random, the probability that it is measured is $\frac{n_2-1}{nk}$ (respectively $\frac{n_2 k-1}{nk}$ for an entire training example).

Finally, if Bob performs measurements with basis $\{|0\>,|1\>\}$, the probability to be detected is $\frac{1}{2}$. This completes the proof.

\subsection{Proof of Theorem \ref{thm:Correctness}}
Since the original perceptron algorithm terminates, there exists a linear classifier $\vec{w}^* = (w^*_{1},w^*_{2},\cdots,w^*_{k})$ and $b^*$, which correctly separates the training set \cite{FreundS1999,Novikoff1963,Rosenblatt1957}.

We first introduce some notations:
\begin{itemize}
  \item $\omega^* = (w^*_{1},w^*_{2},\cdots,w^*_{k},b^*)$, where $\vec{w}^* = (w^*_{1},w^*_{2},\cdots,w^*_{k})$.
  \item $\omega_t = (w_{t,1},w_{t,2},\cdots,w_{t,k},b_t)$, where $\vec{w}_t=(w_{t,1},w_{t,2},\cdots,w_{t,k})$ and $b_t$ denote the $\vec{w}$ and $b$  after $t$ updates, i.e., Eq. \eqref{Eq:UpdateH} is applied $t$ times.
  \item $\chi_t = (x_{i_t,1},x_{i_t,2},\cdots,x_{i_t,k},1)$, where $\vec{x}_{i_t} = (x_{i_t,1},x_{i_t,2},\cdots,x_{i_t,k})$ is the originally training example used in Step \textbf{S2} for $t$-th update. Correspondingly, other variables $\vec{r}$, $c_{i\oplus u}$ and $d$ in Eq. \eqref{Eq:UpdateH} for $t$-th update are represented similarly by $\eta_t$, $y_t$, $\tau_t$ respectively.
  \item for simplicity, denote $\chi'_t = (y_t-\tau_t)\chi_t$ and $\eta'_t = (y_t-\tau_t)\eta_t$. Note $\chi_t\cdot \chi_t = \chi'_t\cdot \chi'_t$ and $\eta\cdot\eta=\eta'_t\cdot \eta'_t$.
  \item Obviously, $\omega_0=0$ by initialization.
\end{itemize}
As the size $N$ of training set is finite, we can further assume that $\forall i, |\vec{w}^*\cdot \vec{x}_i+b^*|>0.$ Then there exists $\gamma>0$ and $R>0$, such that
\begin{equation}\label{Eq:CorrectnessAssumption}
    |\omega^*\cdot\chi_t|>\gamma,  \|\chi_t\|<R, \forall t.
\end{equation}
The above conditions are just the sufficient condition of the termination for the original perceptron algorithm \cite{Novikoff1963}.
Moreover, Eq. \eqref{Eq:UpdateH} can be rewritten as
\begin{equation}\label{Eq:UpdateT}
    \omega_t = \omega_{t-1}+(y_t-\tau_t)\chi_t+(y_t-\tau_t)\eta_t = \omega_{t-1}+\chi'_t+\eta'_t.
\end{equation}
Then this proof works in a similar way as the original proof \cite{Novikoff1963}. That is we also employ Eq. \eqref{Eq:WtBound} to bound $t$.
\begin{equation}\label{Eq:WtBound}
    \|\omega^*\cdot\omega_t\|\leq \|\omega^*\|\|\omega_t\|, \forall t.
\end{equation}

\vspace{10pt}

(1) For the left side, we have
\begin{align*}
    \omega^*\cdot\omega_t &= \omega^*\cdot\omega_{t-1}+(y_t-\tau_t)\omega^*\cdot\chi_t+\omega^*\cdot\eta'_t\\
    &> \omega^*\cdot\omega_{t-1}+\gamma+\omega^*\cdot\eta'_t~~~~~~~~~~~(\text{as~} (y_t-\tau_t)\omega^*\cdot\chi_t>\gamma)\\
    &> \omega^*\cdot\omega_{0}+t\gamma+\sum_{i=1}^t\omega^*\cdot\eta'_i\\
    &= t\gamma+\sum_{i=1}^t\omega^*\cdot\eta'_i.  ~~~~~~~~~~~~~~~~~~~~~~~(\text{as~} \omega_0=0)
\end{align*}
In the above equations, the fact $(y_t-\tau_t)\omega^*\cdot\chi_t>\gamma$ is because
\begin{itemize}
  \item If $\omega^*\cdot\chi_t>0$, we have $y_t=1$. Since $\chi_t$ is employed for updating, it is incorrectly classified by $\omega_{t-1}$, i.e., $\tau_t=0$. Thus by  assumption, Eq. \eqref{Eq:CorrectnessAssumption}, we have $(y_t-\tau_t)\omega^*\cdot\chi_t>\gamma$.
  \item If $\omega^*\cdot\chi_t<0$, similarly $y_t=0$, $\tau_t = 1$ and thus $(y_t-\tau_t)\omega^*\cdot\chi_t>\gamma$.
\end{itemize}

 \vspace{10pt}

(2) For the right side, we have
\begin{align*}
    \|\omega_t\|^2  =&\ \omega_t\cdot\omega_t\\  =&\ \omega_{t-1}\cdot\omega_{t-1}+\chi_t\cdot\chi_t+\eta'_t\cdot\eta'_t+2\omega_{t-1}\cdot\chi'_t+2\omega_{t-1}\cdot\eta'_t+2\chi'_t\cdot\eta'_t\\
    <&\ \omega_{t-1}\cdot\omega_{t-1}+\chi_t\cdot\chi_t+\eta'_t\cdot\eta'_t+2\omega_{t-1}\cdot\eta'_t+2\chi'_t\cdot\eta'_t~~~~~~~~~~~(\text{as~} \omega_{t-1}\cdot\chi'_t<0)\\
    %&~~~~~~~~~~~~~~~~~~~~~~~~~~~~~~~~~~~~~~~~(\text{as~} (y_t-\tau_t)\omega_{t-1}\cdot\chi_t<0)\\
    =&\ \omega_{t-1}\cdot\omega_{t-1}+\chi_t\cdot\chi_t+\eta'_t\cdot\eta'_t+2\omega_{t-2}\cdot\eta'_t+2\chi'_{t-1}\cdot\eta'_t+2\eta'_{t-1}\cdot\eta'_t+2\chi'_t\cdot\eta'_t\\
    <&\ \omega_{t-2}\cdot\omega_{t-2}+\chi_{t-1}\cdot\chi_{t-1}+\eta'_{t-1}\cdot\eta'_{t-1}+\chi_t\cdot\chi_t+\eta'_t\cdot\eta'_t\\
     &+2\omega_{t-2}\cdot(\eta'_{t-1}+\eta'_t)+2\chi'_{t-1}\cdot(\eta'_{t-1}+\eta'_t)+2\eta'_{t-1}\cdot\eta'_t+2\chi'_t\cdot\eta'_t\\
    <&\ \cdots\\
    <&\ \sum_{i=1}^t\chi_i\cdot\chi_i+(\sum_{i=1}^t\eta'_i)\cdot(\sum_{i=1}^t\eta'_i)+2\sum_{i=1}^t\chi'_i\cdot(\sum_{j=i}^t\eta'_j)\\
    <&\ tR^2+(\sum_{i=1}^t\eta'_i)\cdot(\sum_{i=1}^t\eta'_i)+2\sum_{i=1}^t\chi'_i\cdot(\sum_{j=i}^t\eta'_j). ~~~~~~~~~~~~~~~~(\text{as~} \|\chi_i\|<R)
\end{align*}

\vspace{10pt}

(3) The third step is to analyse the randomness from both sides. This is based on the following lemma.
\begin{lem}\label{lem:RandomNumber}
    Suppose $\{X_j\}$ is a sequence of independent identically distributed random variables with the same distribution as $X$ with zero mean $\mathrm{E}X=0$. Suppose $\forall i\in \mathds{N}$, $a_i\in[-R_0,R_0]$ and $\lambda_i\in\{-1,1\}$ only depend on $X_1,\cdots,X_{i-1}$, where $R_0$ is a positive constant real number. For all $t$, define random variables
    $$Y_t = \frac{1}{t}\sum_{j=1}^t\lambda_j X_j,~~~~~~~Z_t = \frac{1}{t^2}\sum_{i=1}^t a_i(\sum_{j=i}^t \lambda_j X_j) = \frac{1}{t^2}\sum_{j=1}^t\lambda_j(\sum_{i=1}^j a_i) X_j.$$
    Then for any $\epsilon>0$, we have $$\lim_{t\ra \infty}\Pr(\|Y_t\|>\epsilon)=\lim_{t\ra \infty}\Pr(\|Z_t\|>\epsilon)=0.$$
\end{lem}
The two results in this lemma can be both proved in a way similar to the weak law of large numbers and the central limit theorem \cite{VanderVaart2000asymptotic,Williams1991probability}, and can be seen as typical excises of characteristic functions in text books, for instance, \cite{VanderVaart2000asymptotic,Williams1991probability}.

\vspace{10pt}

(4) Now we can complete the proof by contradiction. Suppose the algorithm terminates with probability less than $1-\varepsilon$ with $\varepsilon>0$. Then $t$ can go to infinity with probability $\varepsilon$. Since the attribute of $\eta_i$ is independent of each other, by Lemm \ref{lem:RandomNumber}, we have for any $\epsilon>0$,
$$\lim_{t\ra \infty}\Pr(\|\sum_{i=1}^t\eta'_i\|/t>\epsilon)=0,~~ ~~~\lim_{t\ra \infty}\Pr(|\sum_{i=1}^t\chi'_i\cdot(\sum_{j=i}^t\eta'_j)|/t^2>\epsilon)=0.$$

Therefore, there exists a number $T$, such that for any $t>T$, we have
\begin{align*}
    \|\omega^*\cdot\omega_t\|& \leq\|\omega^*\|\|\omega_t\|\\
    \Rightarrow&\ t\gamma+\sum_{i=1}^t\omega^*\cdot\eta'_i < \|\omega^*\|\sqrt{tR^2+(\sum_{i=1}^t\eta'_i)\cdot(\sum_{i=1}^t\eta'_i)+2\sum_{i=1}^t\chi'_i\cdot(\sum_{j=i}^t\eta'_j)}\\
    \Rightarrow&\ t\gamma-0.1t\gamma<\|\omega^*\|\sqrt{tR^2+0.01t^2\gamma^2+0.2t^2\gamma^2/\|\omega^*\|^2}\\
    \Rightarrow&\ 0.5t^2\gamma^2<\|\omega^*\|^2tR^2 \Rightarrow t<2\|\omega^*\|^2R^2/\gamma^2,
\end{align*}
holds with probability 1. Puting it in another way, $t$ is bounded with probability 1. This is a contradiction!

\subsubsection{Proof of Lemma \ref{lem:RandomNumber}}
Since the first one is very similar to the weak law of large numbers \cite{VanderVaart2000asymptotic} and is easier than the other, we only give the proof of the second result.

The key concept in this proof is characteristic functions. The characteristic function of random variable $X$ is defined as $\phi_X(\vartheta) = \mathrm{E}e^{\sqrt{-1}\vartheta X}$ \cite{VanderVaart2000asymptotic}. By the basic properties of characteristic functions \cite{VanderVaart2000asymptotic}, we have
\begin{equation*}
    \phi_{Z_t}(\theta)  = \Pi_{j=1}^t\phi_{\frac{1}{t^2}\lambda_j(\sum_{i=1}^j a_i) X_j}(\theta).
\end{equation*}
Since $X_j$ is independent of  $\frac{1}{t^2}\lambda_j(\sum_{i=1}^j a_i)$, by \cite[Theorem 3]{Wichura2001}, we have
\begin{equation*}
    \phi_{\frac{1}{t^2}\lambda_j(\sum_{i=1}^j a_i) X_j}(\theta) = \phi_{ X_j}(\frac{1}{t^2}\lambda_j(\sum_{i=1}^j a_i)\theta).
\end{equation*}
Note, for any $j<t$,
\begin{equation*}
    \frac{1}{t^2}\lambda_j(\sum_{i=1}^j a_i)\theta = O(\frac{1}{t}), ~~~~~~~~~~~~~\text{as}~~~~~~~~~ t\ra\infty.
\end{equation*}
By \cite{VanderVaart2000asymptotic,Williams1991probability}, we have if $\mathrm{E}X=0$,
\begin{equation*}
    \phi_X(\vartheta) = 1+o(\vartheta), ~~~\text{as}~~~\vartheta\ra 0.
\end{equation*}
Thus for all $\theta$,
\begin{align*}
    \phi_{Z_t}(\theta)  &= \Pi_{j=1}^t\phi_{ X_j}(\frac{1}{t^2}\lambda_j(\sum_{i=1}^j a_i)\theta)\\
    &= \Pi_{j=1}^t(1+o(\frac{1}{t^2}\lambda_j(\sum_{i=1}^j a_i)\theta)) ~~~~~~\text{as}~~~~   t\ra \infty\\
    &= \Pi_{j=1}^t(1+o(\frac{1}{t})) ~~~~~~~~~~~~~~~~~~~~~\text{as}~~~~   t\ra \infty\\
    &\ra 1.  ~~~~~~~~~~~~~~~~~~~~~~~~~~~~~~~~~~~~~~~\text{as}~~~~   t\ra \infty
\end{align*}
This means the distribution of $Z_t$ trends to be the distribution of constant 0, since $\phi(\theta) = 1$ is the characteristic function of constant 0 \cite{VanderVaart2000asymptotic}. Then by L\'{e}vy's theorem \cite{Williams1991probability} and \cite[Theorem 2.7]{VanderVaart2000asymptotic}, we complete the proof.

\section{Detailed Description of Quantum Privacy-Preserving Perceptron Learning Algorithm}\label{Apd:AlgorithmPerceptron}
In this appendix, we present a detailed description of our quantum privacy-preserving perceptron learning algorithm in Algorithm \ref{alg:ProtocolPerceptron}. Since noises is added, this algorithm may not terminate. So we set an upper bound $T$ of loops.
\begin{algorithm}[ht]
    \SetKwData{Left}{left}\SetKwData{This}{this}\SetKwData{Up}{up}
    \SetKwFunction{Union}{Union}\SetKwFunction{FindCompress}{FindCompress}
    \SetKwInOut{Input}{input}\SetKwInOut{Output}{Output}\SetKwInOut{Parameter}{Parameters}

    \Parameter{Training set $D$,\\ the size $N$ of $D$, \\ the upper bound $T$ for the number of loops. \tcp{In the numerical experiments in this paper, $T$ is set to be 40000.}}
    \Output{$\vec{w}$, $b$}
    \Begin{
    $\vec{w}\la 0$, $b\la 0$, $\Delta\la 0$\;
    \For{$t=1,\cdots,T$}{
        $\Delta\la 0$\;
        Alice privately generates $u\in\{0,1\}^n$\;
        \For{$i=1,\cdots,N$}{
            Bob sends $i$ to Alice to query her data system\;
            Alice generates $\vec{r}$ by her private random number generater\;
            Alice sends $c'=c_{i\oplus u}$ and $\vec{x}'=\vec{x}_{i\oplus u}+\vec{r}$ to Bob\;
            \tcp{Step \textbf{S1}}
            Alice runs her quantum data system (Algorithm \ref{alg:ProtocolComputeF}) with input $i\oplus u$ to answer Bob\;\label{line:P6}
            Bob stores result $f(\vec{x}_{i\oplus u})$ in $d$\;
            \tcp{Step \textbf{S2}}
            \If{$d\neq c'$}{\label{line:P8}
                $\Delta\la \Delta+1$\;
                $\vec{w} \la \vec{w}+(c'-d)\vec{x}'$\;
                $b\la b+(c'-d)$\;
            }\label{line:P9}
        }
        \If{$\Delta==0$}{
            Break\;
        }
    \Return{$\vec{w}$, $b$}\;
    }
    }
    \caption{Quantum Privacy-Preserving Perceptron.}\label{alg:ProtocolPerceptron}
\end{algorithm}

\section{Discussion on Classical Reconstruction Method}\label{Apd:Reconstruction}
In Section \ref{Sec:PrivacyAlice}, we mentioned that classical reconstruction method \cite{AgrawalS2000} is not suitable for our problem.
The reason is that since the method \cite{AgrawalS2000} reconstructs the original distribution for each attribute one by one, it does not work correctly when the attributes are strongly related. For instance, each training vector $\vec{x}_i$ in the three sets in Example \ref{Exam:1} has two attributes $\vec{x}_i =(x_{i,1},x_{i,2})$. %This method first estimates the original distribution $h^0_1(a)$ of first attribute of class 0 ($y_i=0$) based on the distribution of $x'_{i,1}=x_{i,1}+r_{i,1}$ and the distribution of $r_{i,1}$.
This method first estimates the probability density function $h^0_1$ for the first attribute of class 0. And then it estimates $h^0_2$, $h^1_1$, $h^0_2$ similarly. When estimating $h^0_1$, it first computes
$$\tilde{h}^0_1(I)= \frac{1}{N}|\{i: y_i=0, x'_{i,1}=x_{i,1} + r_{i,1}\in I\}|,$$
where $I$ is an one-dimensional interval, and then estimates $h^0_1$ based on $\tilde{h}^0_1$ and the probability  density function of noise $r_{i,1}$. Therefore if the value $x_{i,1}$ strongly depends on the value $x_{i,2}$, the estimated $h^0_1$ might be very different to the correct one.

Directly following \cite{AgrawalS2000}, the original method can be modified to estimate the distribution of two attributes together. First we can compute
$$\tilde{h}^0(H)= \frac{1}{N}|\{i: y_i=0, \vec{x}'_{i}\in H\}|,$$
where $H=I_1\times I_2$ is a two-dimensional hypercube, and then estimate $h^0$ based on  $\tilde{h}^0$ and the distribution of noise.

\begin{figure}
  \centering
  \subfigure[1D reconstruction, training set 1.]{\includegraphics[width=5cm]{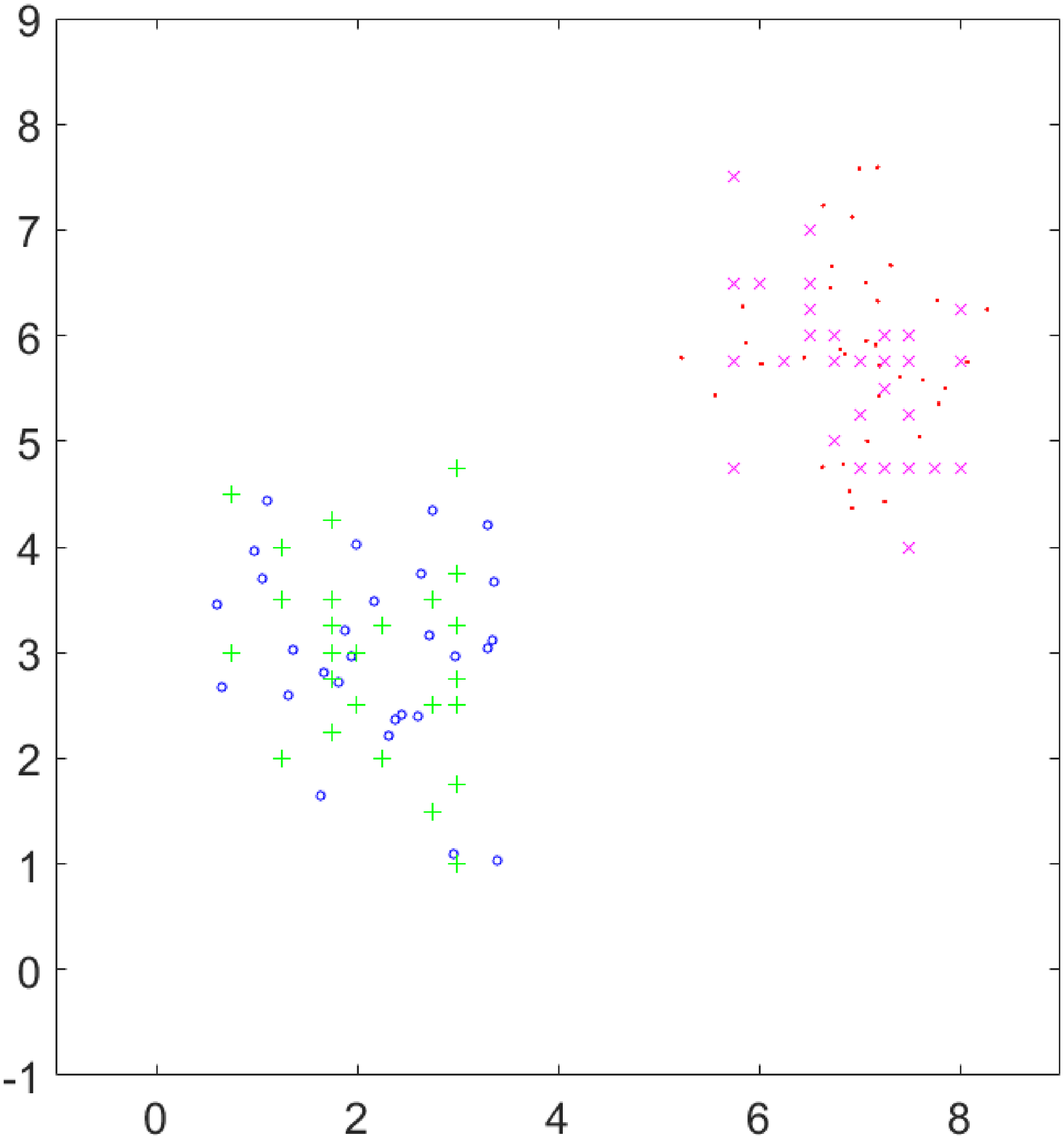}}
  \subfigure[2D reconstruction,  training set 1.]{\includegraphics[width=5cm]{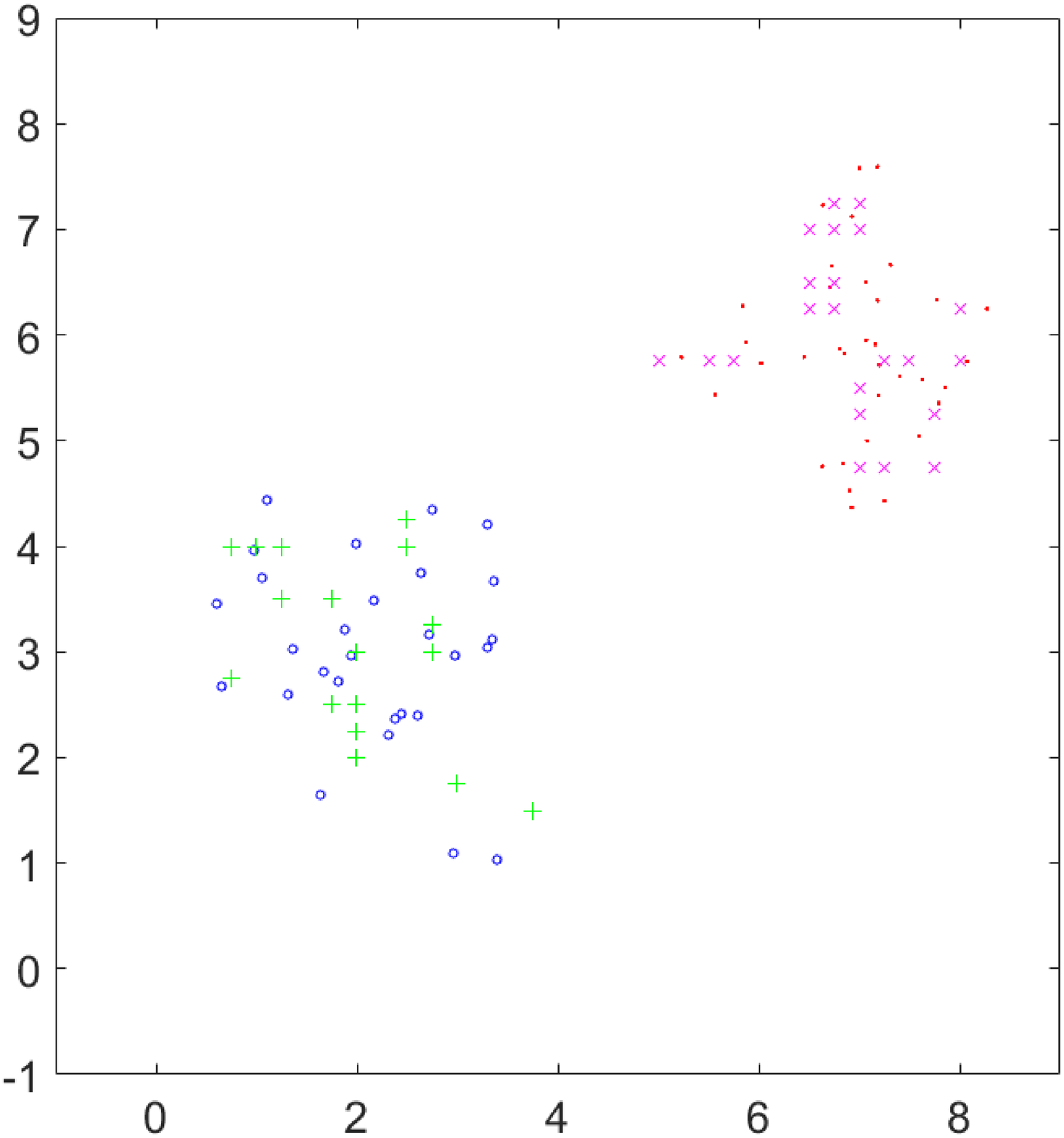}}
  \subfigure[1D reconstruction, training set 2.]{\includegraphics[width=5cm]{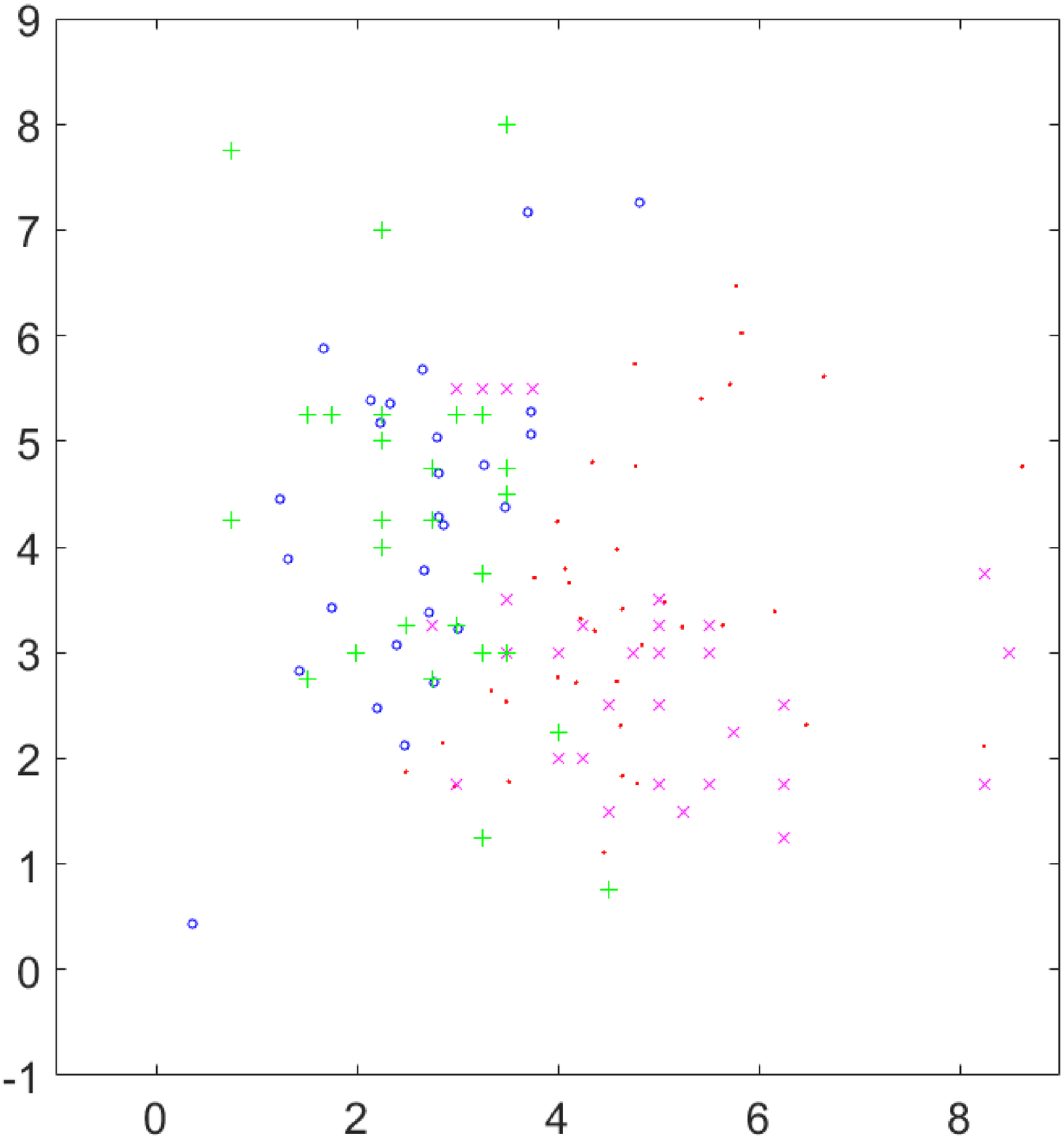}}
  \subfigure[2D reconstruction,  training set 2.]{\includegraphics[width=5cm]{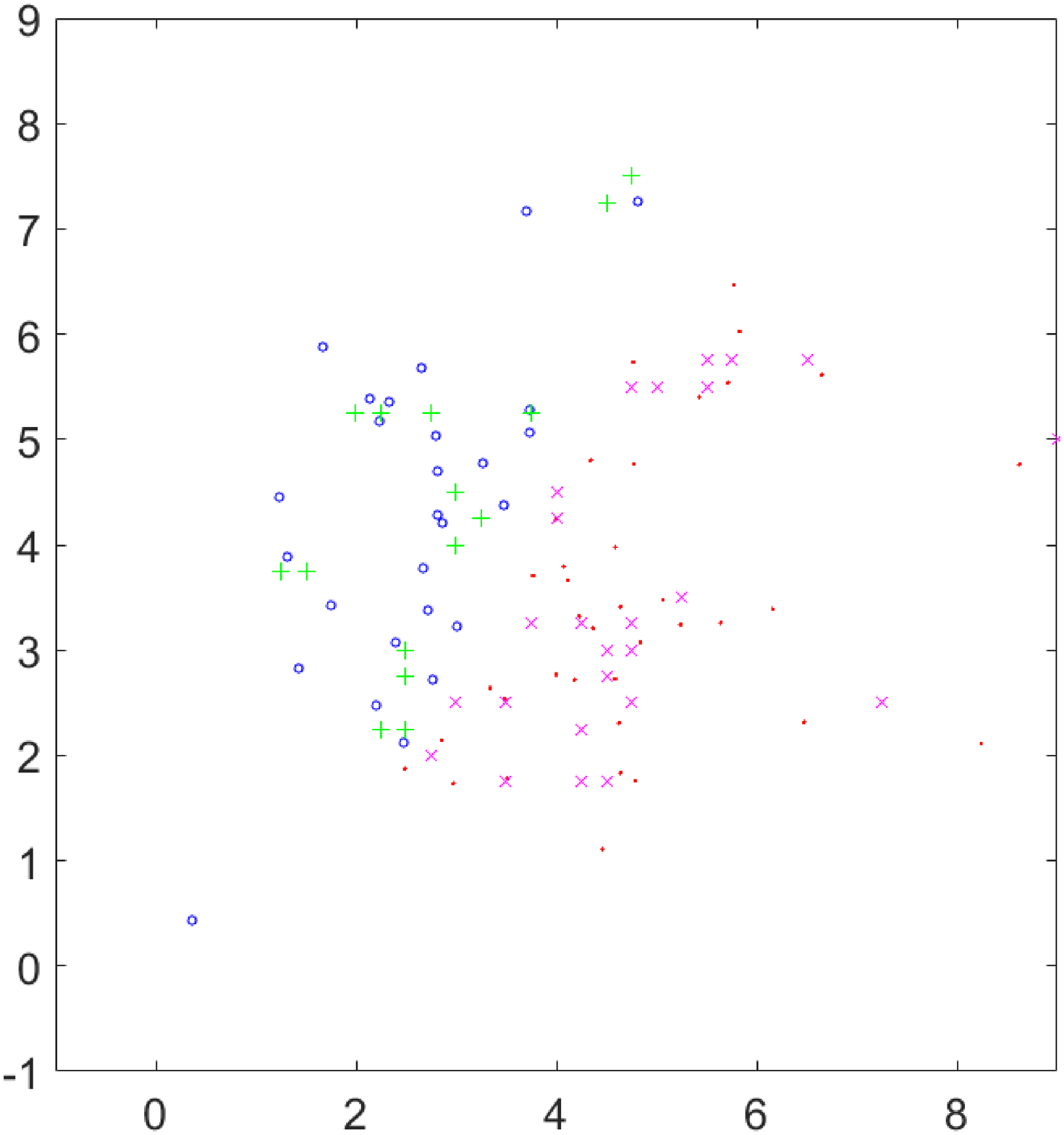}}
  \subfigure[1D reconstruction,  training set 3.]{\includegraphics[width=5cm]{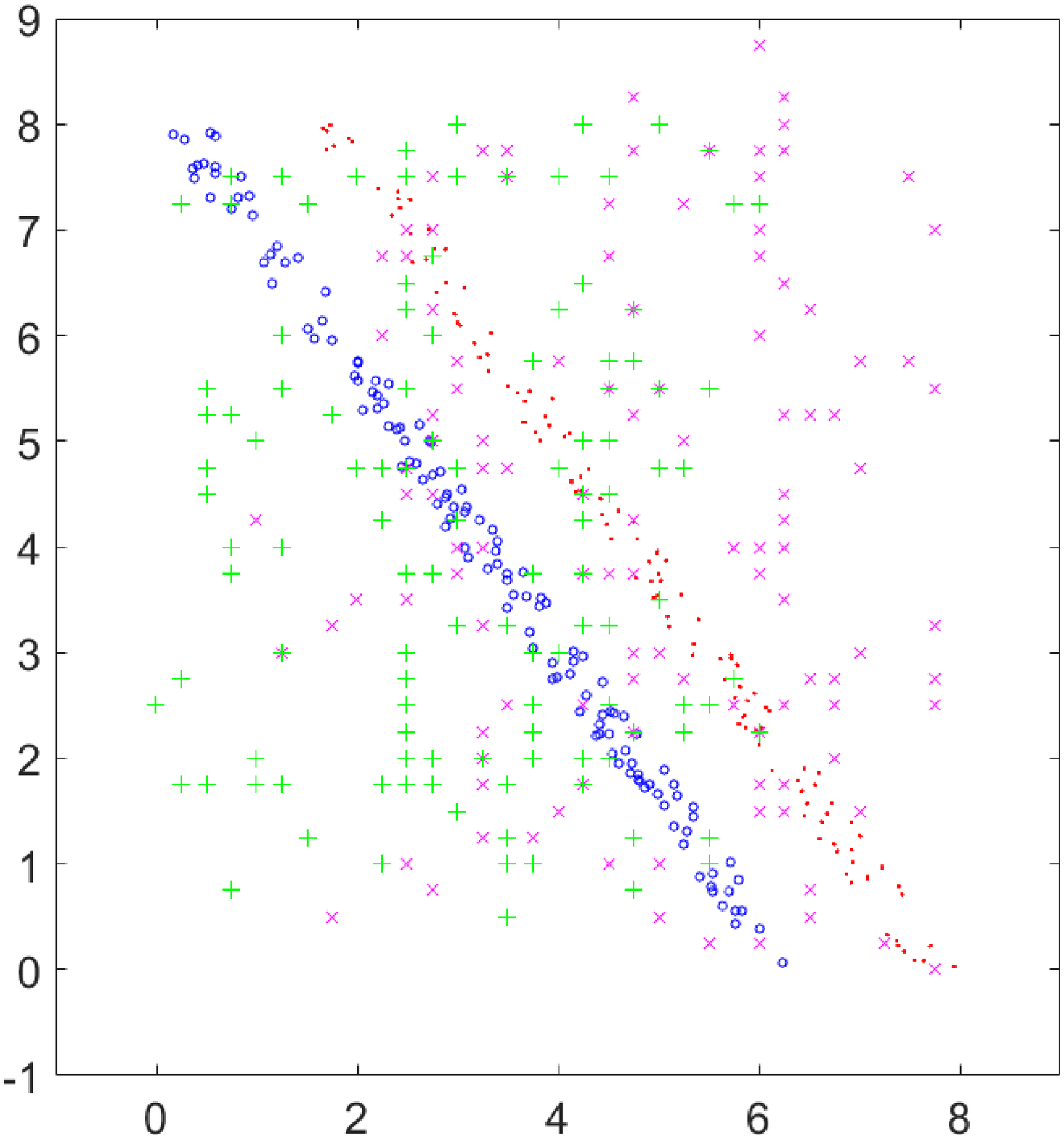}}
  \subfigure[2D reconstruction,  training set 3.]{\includegraphics[width=5cm]{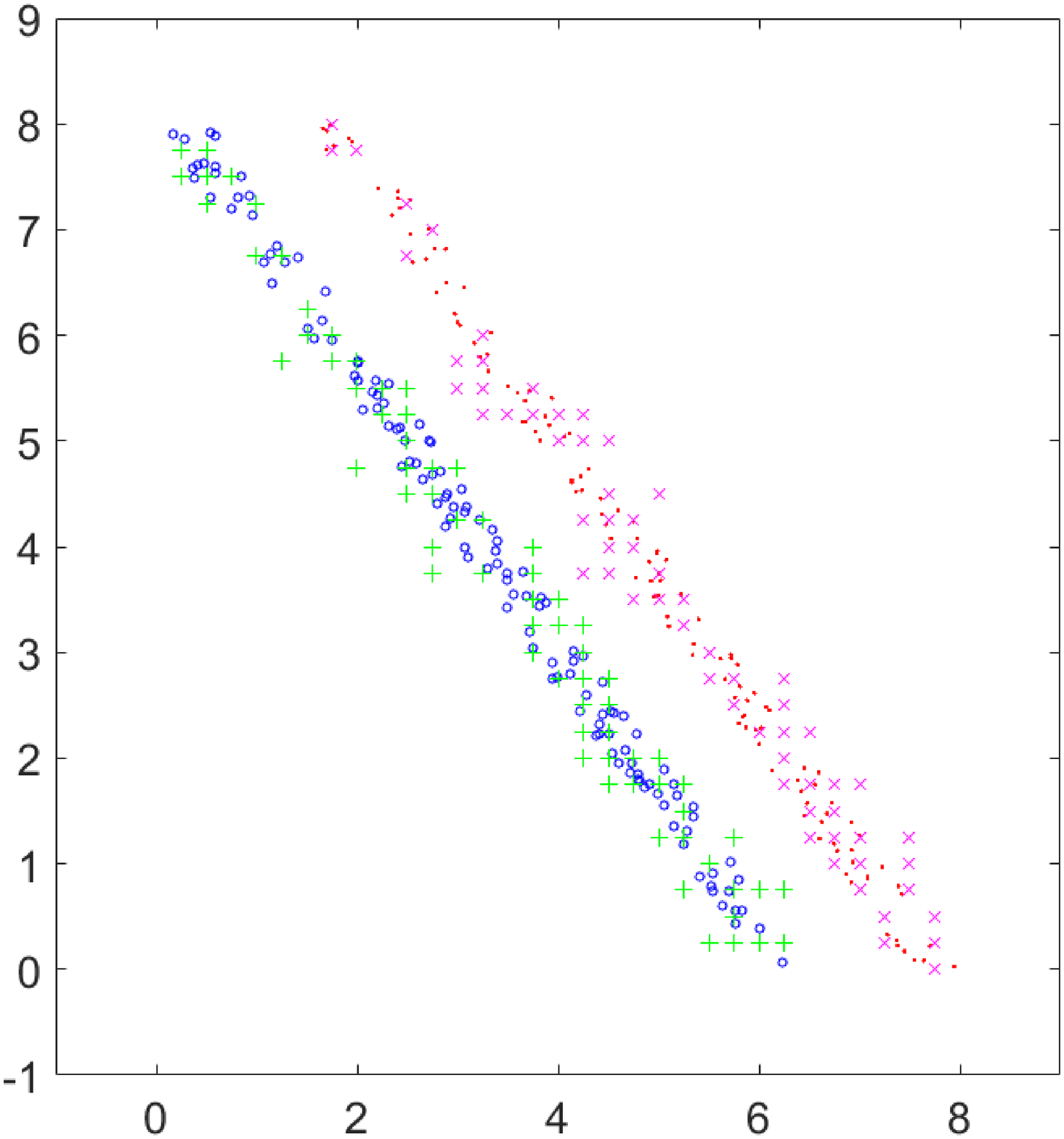}}
  \caption{Reconstructed training sets in Example \ref{Exam:1}.}\label{Fig:Exam1Recon}
\end{figure}

We give a direct view of the results of these two different methods on the training set in Example \ref{Exam:1}. The reconstructed training sets (for only one execution, not averaged) is shown in Fig. \ref{Fig:Exam1Recon}.

Obviously this modified two-dimensional method works better than the original one-dimensional one. But unfortunately, based on the analysis in \cite{AgrawalS2000}, its time complexity is $O(NL^k)$, which is exponential on $L$, where $L$ is the number of one-dimensional intervals that one axis is divided into. For instance, if the range of $x_{i,1}$ is [-16,16] and we divides it into $L=20$ intervals, then the total number of hypercubes for training vectors $\vec{x}$ is $L^2 = 400$.

\end{document}